\documentclass[journal, a4paper]{IEEEtran}
\usepackage{stackengine}
\usepackage{amsmath}
\usepackage{scrextend}
\def\yenrule{\rule{1.3ex}{.1ex}}
\usepackage[Algorithm]{algorithm}
\usepackage{algpseudocode}
\usepackage{textcomp}
\usepackage[australian]{babel}
\usepackage{ wasysym}
\usepackage{setspace}
\usepackage{graphicx}
\usepackage{multicol}
\usepackage{multirow}
\usepackage{rotating}
\usepackage[table]{xcolor} 
\usepackage{booktabs}
\usepackage{subfloat}
\usepackage{subcaption}
\usepackage{xcite}
\usepackage{mathtools}
\usepackage{comment}
\usepackage{url}
\usepackage{authblk}
\def\textyen{\renewcommand\stacktype{L}\stackon[.4ex]{\stackon[.65ex]{Y}{\yenrule}}{\yenrule}}
\graphicspath{{figures/}}

\providecommand{\keywords}[1]{\textbf{\textit{Index terms---}} #1}

\begin{document}

\title{Scheduling Algorithms for Efficient Execution of Stream Workflow Applications in Multicloud Environments}

\author[1]{Mutaz Barika}
\author[1]{Saurabh Garg}
\author[1]{Andrew Chan}
\author[2]{Rodrigo N. Calheiros}
\affil[1]{University of Tasmania, Australia}
\affil[2]{Western Sydney University, Australia}

\maketitle

\begin{abstract}
	Big data processing applications are becoming more and more complex. They are no more monolithic in nature but instead they are composed of decoupled analytical processes in the form of a workflow. One type of such workflow applications is stream workflow application, which integrates multiple streaming big data applications to support decision making. Each analytical component of these applications runs continuously and processes data streams whose velocity will depend on several factors such as network bandwidth and processing rate of parent analytical component. As a consequence, the execution of these applications on cloud environments requires advanced scheduling techniques that adhere to end user's requirements in terms of data processing and deadline for decision making. \textcolor{black}{In this paper, we propose two Multicloud scheduling and resource allocation techniques for efficient execution of stream workflow applications on Multicloud environments while adhering to workflow application and user performance requirements and reducing execution cost.} Results showed that the proposed genetic algorithm is an adequate and effective for all experiments.
\end{abstract}


\section{Introduction}\label{sec:introduction}

\textcolor{black}{The continuous evolution of Internet of Things (IoT) and its fast adoption are driving change in data ecosystems. Beyond the hype and in near reality, i.e. by 2020, there will be tens of billions of IoT devices\cite{hassan2017internet} and all of these devices generate data, leading to exponential data growth. On the one hand, this imposes new data processing challenges, but on the other hand, it opens up opportunities for designing and developing applications and services that utilize IoT data to facilitate real-time data analysis for online insights.}

\textcolor{black}{Recently, a number of streaming data processing platforms have been developed to transact with data streams being generated with great velocity, which allows designing and building streaming big data applications to ingest, process, and analyse data streams. However, the need of composing these applications into data pipelines to make better decisions in real-time is increasingly required. Following this need, many IoT applications and services such as smart traffic control, smart irrigation and forest fire detection, are evolving to cope with the demand of improving the quality of our lives \cite{zanella2014internet} \cite{mehmood2017internet} \cite{bahga2014internet}. These applications are not monolithic application but they reys on the technology of workflow to model different analytical components, where each component can be a simple analytical step or a workflow itself,  to make better decisions. An example of this workflow is smart road traffic monitoring as a service of smart city services that utilizes processing power of connected vehicles in addition of roadside infrastructure (e.g. traffic lights, cameras) to create real-time view of road traffic conditions [3]. This type of workflow is also called stream workflow.}

\textcolor{black}{In contrast to business and scientific workflows which are static workflows, stream workflow supports composition of analytics components into a holistic data processing pipeline to perform complex and continuous data computation operations over infinite data streams with great velocity. Stream workflows are very different from traditional business and scientific workflows as these workflows have to continuously process an infinite stream of data with each analytical component always in an active state. Moreover, each component has heterogeneous platform and infrastructure requirements. Furthermore, stream workflow differ from streaming operator graphs (generated by streaming data platforms) as there is a single source of data for the whole operator graph and one end operator, while stream workflow has multiple input data sources and multiple output streams. Therefore, the complexity and heterogeneity of stream workflow applications and cloud compute resources, the distribution of external data sources for these applications, and user-defined quality of service (QoS) requirements, all impose the need for a new class of application orchestration workflow. This orchestration process of such application in the cloud includes managing pipeline execution dependencies, scheduling and concurrency based on application data flow.}

\textcolor{black}{Unfortunately, existing research works for big data processing have focused on supporting batch-oriented workflows, providing streaming operators graph for continuous processing or developing big data orchestrators that do not need to guarantee real-time data processing requirements. Research works like \cite{wang2009kepler+} and \cite{wang2014big} use a batch processing model to provide architectures to compose batch processing applications into a pipeline to process static data at once and get final analytical insights by extending the capability of scientific workflow management systems such as Kepler \cite{altintas2004kepler} and Pegasus \cite{deelman2015pegasus}. Systems like Apache Storm \footnote{\textcolor{black}{https://storm.apache.org/}} and Flink \footnote{\textcolor{black}{https://flink.apache.org/}} employ continuous operator model to process data streams. However, these systems form stream operators graph with one feeding data source and one sink operator to output the result of graph execution, while stream workflow has multiple input data sources and multiple output streams. Moreover, they concentrate on minimizing the latency and fail to maintain high throughput. Big data orchestrators (Apache YARN, Apache Mesos) provides script-based composition of analytical steps over cloud datacentres. However, these orchestrators assume either monolithic applications that do not need to meet real-time decision support requirements defined by workflow owner or are intended for workflows that have predictable performance \cite{ranjan2017orchestrating}.}

\textcolor{black}{As stream workflow is different from the models discussed above, research works and systems in the literature looked at the composition of streaming applications from different perspectives of scheduling problem. Moreover, the execution of stream workflow on resources provisioned from single cloud may not meet user requirements due to the distribution of external data sources. Multicloud environment consolidates multiple clouds, allowing to orchestrate the execution of multiple analytical components over different clouds to utilize data locality. However, the composition needs of stream workflows and the dynamic nature of cloud computing poses a challenge in the problem of executing such workflow over different cloud infrastructures efficiently while meeting user real-time user requirements.}

\textcolor{black}{In this paper, we address the challenge of determining near-optimal resource allocation and scheduling of stream workflow applications to meet user requirements while reducing the total cost of execution with the use of Multicloud architecture. To this end, we design and implement two efficient scheduling algorithms using Greedy and Genetic heuristics. We evaluate their efficiency by comparing them using commonly types of real workflow structures in different experiment scenarios and present experimental results. Our contributions are as follows:}

\begin{itemize}
	\item Problem modelling of stream workflow application.
	\item A greedy resource provisioning and scheduling algorithm for efficient execution of stream workflow.
	\item A genetic resource provisioning and scheduling algorithm for efficient execution of stream workflow.
	\item A comprehensive analysis of the two proposed algorithms (greedy and genetic algorithms) using various structures of workflow with different sizes.
\end{itemize}

\section{Related Work}

\textcolor{black}{Given the complexity and heterogeneity of stream workflows and the compute resources in addition to user-defined quality of service (QoS) requirements, it represents a new class of scheduling problem. The orchestration process of stream workflow application over cloud infrastructures is not a trivial task. However, most of research works focused on big data batch computing and thus big data stream computing is still receiving little attention. Thus, few scheduling methods found in the literature that are related to our work.} 

\textcolor{black}{Looking at the streaming operator graphs that used with stream processing systems, it is important to determine the differences between these graphs and stream workflows that we consider in this paper, and therefore how our scheduling problem is different. Stream-oriented big data platforms and services such as Apache Storm and Azure Stream Analytics provide the ability to design streaming operator graphs to process streams and produce final output stream. First of all, streaming operator graphs that generated by those systems differ from stream workflows as the source of data for the whole operator graph is one and there is one end operator, while stream workflow has multiple input data sources and multiple output streams. Moreover, each component in stream workflow has heterogeneous platform and infrastructure requirements. Furthermore, the goal of these systems is to attain low stream latency without taking into consideration other optimisation goals such as network usage, execution performance and cost.}

\textcolor{black}{In regards to the most related research works \cite{pietzuch2006network}, \cite{cardellini2016optimal} and \cite{venkataraman2017drizzle}, these works also addressed the placement problem for those operator graphs in distributed large-scale environments with various limitations. Pietzuch et al. \cite{pietzuch2006network} proposed operator placement algorithm (called SBON) that optimizes operator placement to enhance network utilization by using the continuous knowledge of network and node conditions (i.e. network usage metric), aiming at providing low latency. This research work lacks the consideration of the location of data stream sources in making placement decisions. In the same context, Cardellini et al. \cite{cardellini2016optimal} proposed optimal placement model and prototype scheduler for operator graphs that optimised user-oriented QoS attributes. This work only presented the modelling of network-related QoS attributes (elastic energy, network usage and inter-node traffic), and made the consideration of other constraints such as execution cost or performance as future research directions. It also ignores the user-defined performance constraints on the operator graph. Venkataraman et al. \cite{venkataraman2017drizzle} focused on optimizing the scheduling of operator graph and presented techniques that are implemented in Drizzle to enable high throughput and adaptability, and low latency. This work ignores the consideration of data source location, relies on micro-batch processing system (i.e. Apache Spark) to provide stream processing at scale. It also lacks the consideration of user-oriented QoS attributes. Accordingly, the placement problem of operator graph is related to a different type of stream graph application as well as has different assumption and optimization goals in comparison to the stream workflow and the scheduling problem that we consider in this paper. With stream workflow, the scheduling problem considers the mapping of each analytical component to one or more compute resources as well as the optimization goals are minimizing execution cost and improving performance without violating real-time user requirements.}

\textcolor{black}{In regards to big data orchestration, existing big data orchestrators that can be extended for big data management are Apache YARN \cite{vavilapalli2013apache} and Apache Mesos \cite{hindman2011mesos}. Each of these systems uses a different scheduling technique to map applications on cloud resources.}

\par Apache YARN uses a monolithic scheduler (single centralized scheduler) to map compute resources among competing applications in the cluster. Apache YARN was designed for optimizing the scheduling of Hadoop jobs (i.e. batch jobs), but not for services with long runtimes, such as Service Oriented Architecture (SOA) applications, nor for short-lived interactive queries (real-time workloads), such as stream jobs, and ``\emph{while it’s possible to have it schedule other kinds of workloads, this is not an ideal model}'' \cite{scott2015mesosyarn}. It also was not designed for stateful services, instead it is appropriate for stateless batch jobs that can be restarted easily in case of failures \cite{scott2015mesosyarn}. Furthermore, it is designed to support homogeneous clusters of IaaS resources \cite{ranjan2015cross}, and does not support workflows or dynamic composition of data-intensive activities.

\par Apache Mesos uses a dual-level scheduling mechanism called ``resource efforts'', which makes offer of resources (a list of available resources on multiple slave nodes) to a framework and let this framework either accept the offer, or reject it if the offered resources do not meet its constraints and then waits for ones they do \cite{hindman2011mesos}. Therefore, the responsibility of the master is to decide how many resources to offer each framework according to an allocation policy (such as fair sharing or priority) defined by the system administrator via a pluggable allocation module, while frameworks take the responsibility for deciding which offered resources to accept as well as which workloads to run on them \cite{hindman2011mesos}. Mesos is designed for homogeneous clusters of IaaS resource \cite{ranjan2015cross}, and it does not deal with the complexity and dynamism of big data workflows.

\textcolor{black}{The aforementioned big data orchestrators assume either that they do not need to meet real-time decision support requirements \cite{ranjan2017orchestrating} or are intended for big data workflows that have predictable performance \cite{ranjan2017orchestrating}. Thus, the scheduling techniques in those orchestrators are considered the big data workflow application as a static structure, so that they neglect the dynamic nature of this application and its analytical components, the unpredictable performance of this workflow application, real-time performance requirements defined by the owners of these workflows, the runtime changes and the powerful capability of 'cloud of clouds' as an dynamic execution environment. Accordingly, these orchestrators do not fit the composition needs of complex big data workflows. They also do not leverage the capability of Multicloud environment to cope with the dynamic aspects of these workflows.}

\textcolor{black}{In accordance to the above overall discussion and for stream workflow application as a one type of big data workflow application, scheduling and resource allocation technique is needed to execute this application efficiently in multiple cloud infrastructures while meeting user real-time requirements and reducing the execution cost.}

\section{Multicloud Execution Environment and Stream Workflow Application}

\subsection{Overview of Multicloud Environments}
\textcolor{black}{When targeting distributed data sources that inject their data streams into a workflow pipeline, it is necessary to utilize data locality by leverage Multicloud architecture. If all resources are provisioned from a single cloud and not all data sources are near this cloud, transfer of large amounts of data to the corresponding resources not only leads to the difficulty of achieving the requirements of real-time data analysis, but also is expensive and incurs high latency. Moreover, if the location of the data source changes at any time, the flexibility provided by a multicloud architecture allows the corresponding analytical component to be moved to the new data location. Furthermore, if the amount of data produced by a data source decreases overtime and reaches low data rate, the opportunity to move the corresponding analytical component to another cloud helps to improve performance and reduce the cost without violating user real-time requirements. A single cloud cannot deal with all of the aforementioned points, and thus a multicloud architecture should be preferred in these scenarios.}

\begin{figure*}
	\centering
	\includegraphics[width=0.7\textwidth]{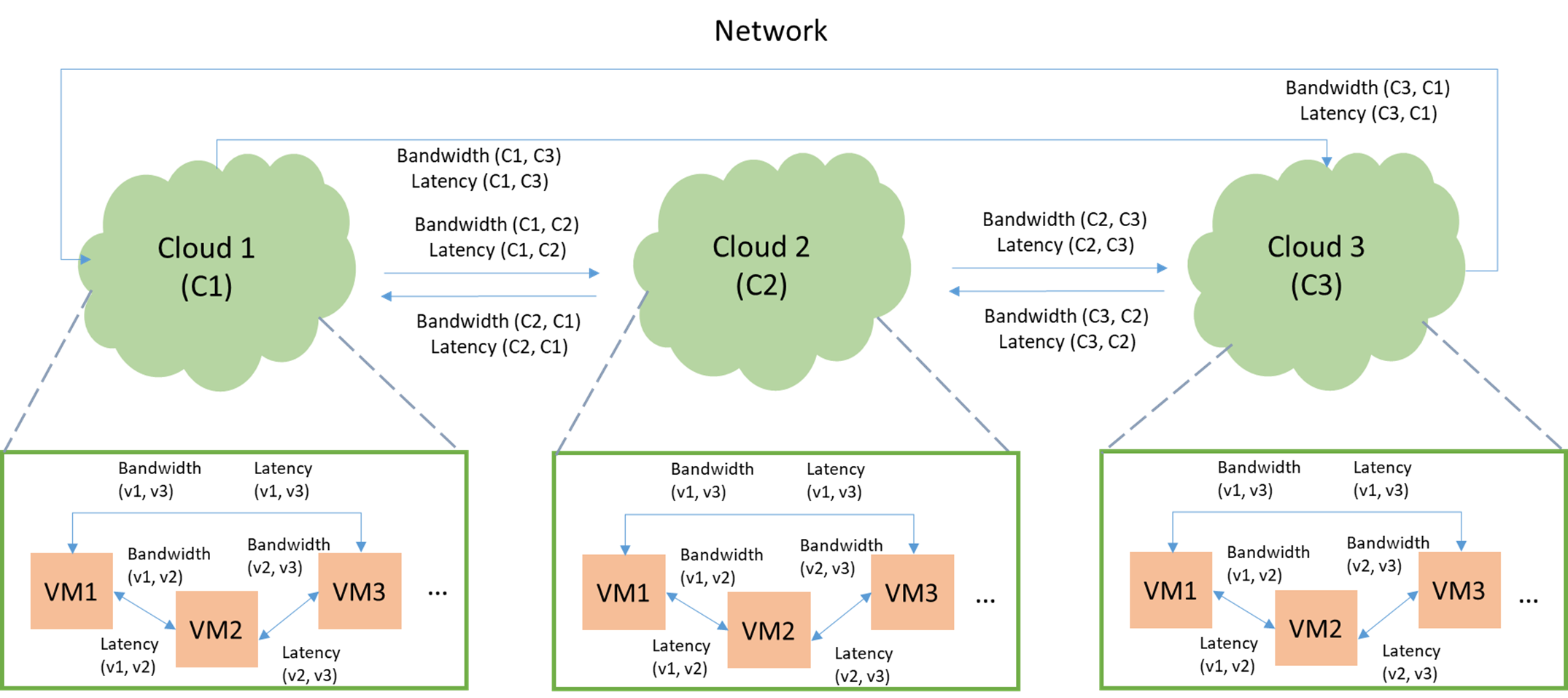}
	\caption{Multicloud environment: network.}
	\label{fig:multicloudenv}
\end{figure*}

\textcolor{black}{A global view of a multicloud environment is depicted in Figure \ref{fig:multicloudenv}. Each cloud is independent from other clouds and offers different levels of compute capacity at different costs. The network bandwidth between compute resources in one cloud is mostly unchanged, while between various clouds is different and variable. Similarly, the latency between compute resources in one cloud is mostly low, while between various clouds can be comparatively high.}

\begin{figure*}
		\centering
		\includegraphics[width=0.6\textwidth]{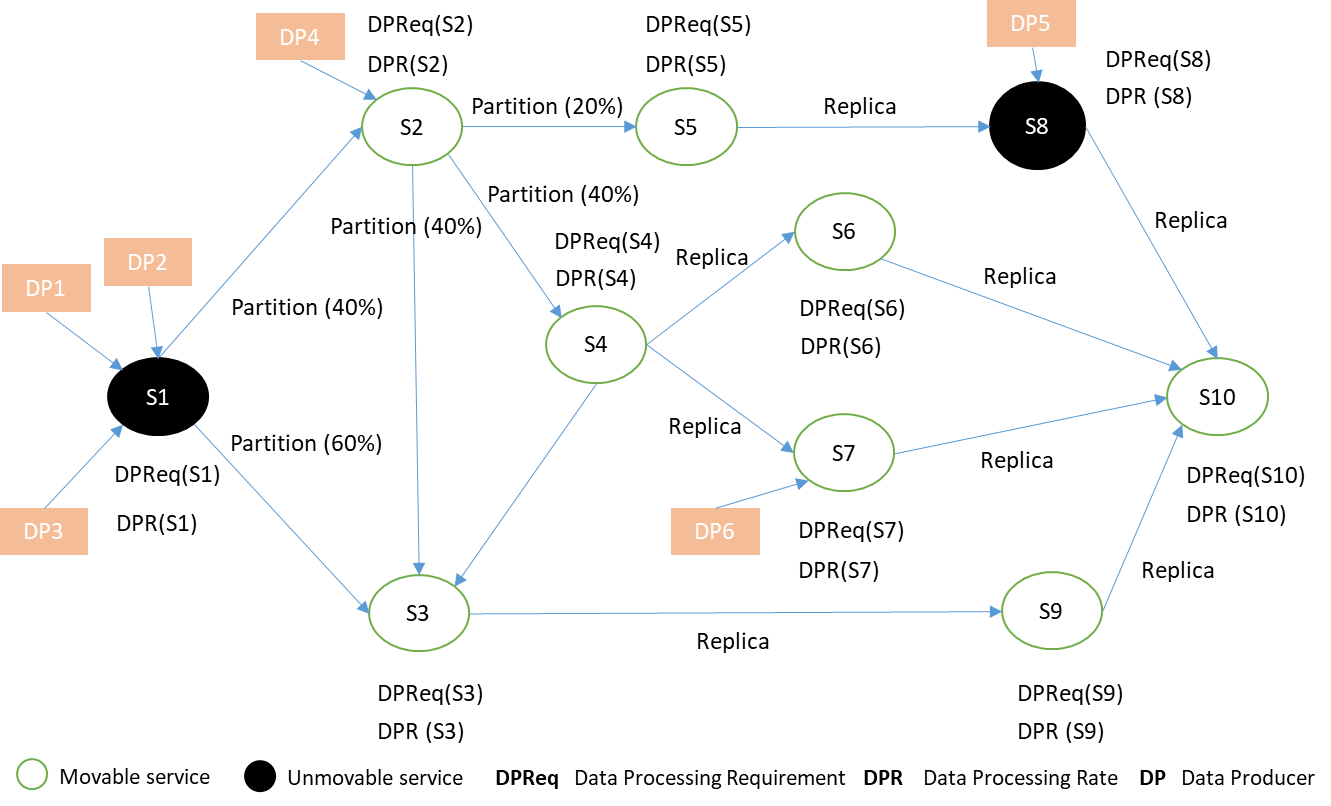}
		\caption{Stream workflow application example.}
		\label{fig:bigdataworkflows}
\end{figure*}

\subsection{Stream Workflow Applications and their Requirements}

\textcolor{black}{Stream workflow applications comprise multiple streaming analytical components, which can be seen as services, as they can independently execute over any virtual resources, although data dependencies among them should be maintained. With this workflow application, we deal with continuous inputs from internal sources (i.e. output data of parent services) as well as from external sources (such as sensors), continuous data processing that is carried out by running services for incoming data and continuous outputs that are results of processing data at services, which routed towards one or more child services. The end services generate the continuous output results for the execution of this workflow. Figure \ref{fig:bigdataworkflows} shows an example of stream workflow application with its requirements. With this workflow application, the two types of services are:}
\begin{itemize}
	\item Unmovable service: It is a service with unmovable data, which means the data volume coming from data stream sources is large and we need to process such data locally to avoid the cost and time of transfer data such as vehicle detection service. Thus, data locality approach is applied with this service; or 
	\item Moveable service: It is a service with movable data, which means the working stream is small and can be transferred with low communication overhead of data transmission. Thus, placement optimization approach is applied to exploit deployment flexibility.
\end{itemize}

\textcolor{black}{As noted in Figure \ref{fig:bigdataworkflows}, each service has its own data processing requirements, which is the number of instructions required to process one MB of stream data, and data processing rate, which is the measure of the amount of data that can be processed in a given time by a service (in MB/s). In term of the mode of data that being routed towards one or more child services, there are two data modes:}
\begin{itemize}
	\item Replica mode: The child service receives replica copy of the output stream of a parent service.
	\item \textcolor{black}{Partition mode: The child service receives a portion of the output stream of a parent service according to the specified partition percentage.}
\end{itemize}

\textcolor{black}{The owner of stream workflow application allows to specify maximum performance constraints in terms of data processing rate of services targeting the maximum desired processing performance that she/he is willing to pay for achieving it during the whole execution, and letting the cost minimization carried-out at initial scheduling plan. If no performance constraints are specified, the initial input data rates of services are considered as the maximum performance constraints (representing maximum desired processing performance for those services). Of course, the input data rate is varying overtime, so that a strategy is needed to handle the increase in data rate. We assume that the exceed incoming data rate will be dropped, thus the increase of load above the pre-specified maximum throughputs will have no effect. Of course, if the speed of incoming data streams decreases, the throughput of service still has the full capability to handle the increase in the speed of data upto the pre specified processing performance. In addition to achieving user specific performance constraints in term of throughputs of services, the end-to-end latency (response time) is crucial in stream workflow application. It is the time between receiving a data stream at a service and generating output stream that regards this stream. Ensuring the low latency is required during the whole execution of stream workflow. It should be kept as low as possible or be bounded when it starts to increase whilst maintaining user specific throughput.}

\textcolor{black}{Accordingly, the variables of stream workflow are service type, its data processing requirement, its data processing rate and the dynamism of execution environment. The latter includes network bandwidth and latency between different clouds. As a result, both user performance requirements and workflow application requirements need to be considered and achieved in addition to maintaining low latency during the execution of this workflow.}

\section{Problem Modelling}
\par Prior to introduce the problem modelling of stream workflow application, we list all the terminologies that will be used in this model in Table \ref{tab:notation}.

\begin{table}
	\scriptsize
	\centering
	\begin{center}
		\caption{\textcolor{black}{Problem Modelling Notation}}
		\begin{tabular}{|p{1.7cm}|p{6.3cm}|}
			\hline 
			Symbol / Term & Description \\ 
			\hline 
			G & Workflow graph \\
			\hline 
			S & Set of all graph services \\
			\hline 
			E & Set of all graph edges \\
			\hline 
			$\textyen^m$ & Percentage of data that is routed from parent service to child service (100\% in replica mode or any percent in partition mode) \\
			\hline 
			$S_n$ & Particular service in workflow graph \\
			\hline 
			$MI^{S_n}$ & Number of floating-point operations required to process one MB of input data (MI/MB)\\
			\hline 
			$\lambda^{S_n}$ & Amount of data produced by a given external source and being consumed by a service (MB/s) \\
			\hline 
			$\gamma^{S_n}$ & Proportion of output data to input data for $S_n$\\
			\hline 
			C & Set of all clouds in Multicloud environment \\
			\hline 
			$c_g$ & Particular Cloud in Multicloud environment\\
			\hline 
			L & Network latency matrix \\
			\hline 
			B & Network bandwidth matrix \\
			\hline 
			D & Data transfer cost matrix \\
			\hline 
			$VM^g$ & Set of all VMs in cloud g \\
			\hline 
			$vm_k^g$ & Particular VM k in cloud g \\
			\hline 
			$U^g$ & Set of all internal network links between VMs in cloud g \\
			\hline 
			$u_h^g$ & Particular internal link between $vm_{org}^g \text{ and } vm_{dest}^g$ \\
			\hline
			$MIPS_{vm_k^g}$ & Rating of the capacity of VM k in cloud g \\
			\hline 
			$\cent_{vm_k^g}$ & Provisioning cost of VM k in cloud g (cents/s) \\
			\hline 
			$\mu^{S_n}$ & user-defined maximum performance constraint (MB/s) \\
			\hline
			$\alpha_{S_n}$ & Data processing rate of $S_n$ \\
			\hline
			unitDUnit & Minimum stream unit for the whole application (MB) \\
			\hline 
			unitDPRate & Minimum stream unit per second for the whole application (MB/s) \\
			\hline 
		\end{tabular}
		\label{tab:notation}
	\end{center}
\end{table}

\subsection{Application Model}
We model a stream workflow application as a Direct Acyclic Graph (DAG) \(G = (S, E)\). S represents a set of N services \(S = {s_1, s_2, ..., s_N}\) and E represents a set of M edges/links between services denoted as \(E = {e_1,e_2, ..., e_M}\). Each edge, \(e_m\) is represented as a tuple \((s_{org}^m, s_{dest}^m, \textyen^m)\), where \(s_{org}^m\) denotes origin service, \(s_{dest}^m\) denotes destination service and \(\textyen^m\) denotes the percentage of data generated by \(s_{org}^m\) that is routed towards \(s_{dest}^m\). \\

Each particular service \(S_n\), is represented as a tuple \(S_n = (MI^{S_n}, \lambda^{S_n}, \gamma^{S_n})\), where \(MI^{S_n}\) denotes the number of floating-point operations required to process one MB of incoming data (service data processing requirement) in MI/MB, \(\lambda^{S_n}\) denotes the arrival rate of data streams generated by sources outside the application in MB/s (such as data streams generated by sensors) to be consumed by the service, and \(\gamma^{S_n}\) denotes the proportion of data generated by a service based on input streams.\\

Notice that, given the nature of stream workflow applications, it is possible that data generated by one service can be sent to one or more services, or can be split among different services. Thus, for service \(S_n\), both parameters \(\gamma^{S_n}\) and \(\textyen^m\) (in edges where such service is origin service) are necessary to define the whole application. In addition, to process streams that coming at different speeds, the minimum stream unit per second (denoted as unitDPRate) is needed to be specified for the whole application, so that each provisioned compute resource must process at least one unit per second and of course it can process multiple units per second according to its computing capacity per second (in term of MIPS). By specifying minimum stream unit per second for the whole workflow application, the data processing rate (MB/s) for processing this unit can be determined to ensure that each provisioned compute resource at least processes one unit per second.

\subsection{System Model}
The cloud system is modelled as a tuple W = (C, L, B, D). A set of G clouds in the Multicloud environment is denoted as \(C = {c_1, c_2, ..., c_G}\). L, B, and D denote matrices containing respectively the latency (in seconds), the bandwidth (in MB/s), and the data transfer cost (in cents/MB or \cent/MB) between each of the pair of clouds in C.

Each cloud, \(c_g\) is represented as a tuple \( (VM^g, U^g) \), where \( VM^g = {vm_1^g, vm_2^g, ..., vm_K^g} \) is a set of K virtual machines (compute resources) with different resource configurations deployed in \(c_g\), and \( U^g = {u_1^g, u_2^g, ..., u_H^g}, u_h^g = (vm_{org}^g, vm_{dest}^g) \), a set of H links that are part of the datacenter network topology.

Each VM deployed in the cloud, \(vm_k^g\), is represented as a tuple \( (MIPS_{vm_k^g}, \cent_{vm_k^g})\), where \(MIPS_{vm_k^g}\) denotes floating-point operations computed by this VM according to its compute capacity per second and \(\cent_{vm_k^g}\) denotes the cost of provisioning such VM (in cents per second).

The data processing rate for \(S_n\) if it is mapped to \(vm_k^g\) is denoted as \(\varphi_k^g\) and is calculated as:
\begin{equation} 
\varphi(S_n, vm_k^g) = \frac{\lfloor MIPS_{vm_k^g} / \chi \rfloor \ast \chi}{MI^{S_n}} \text{  MB/s}
\end{equation}
\[Where \text{     } \chi = unitDPRate \ast MI^{S_n} 
	and \text{     } MIPS_{vm_k^g} \geq \chi
\]

\textcolor{black}{The workflow application owner can specify maximum performance constraint for service \(S_n\) (denoted as \(\mu^{S_n}\)) as a part of request (in MB/s) as a value for data processing rate of service \(S_n\) (denoted as \(\alpha_{S_n}\)), targeting the maximum desired processing performance that she/he is willing to pay for achieving it during the whole execution. If no performance constraint for service \(S_n\) is specified, the system will calculate this rate based on input stream(s) of service \(S_n\). In that case, each service \(S_n\) is capable to handle upto the specified data processing rate (throughput) and when the speed of input streams increases this maximum throughput \(\mu^{S_n}\), the dropping mechanism is applied. Of course, if the speed of incoming data streams decreases, \(S_n\) still has the full capability to handle the increase in the speed of data upto \(\mu^{S_n}\).} Let \(pro(S_n)\) be a set of VMs
that are provisioned from one cloud for service \(S_n\) and \(inStream(S_n)\) denote the input stream of \(S_n\).

The \(inStream(S_n)\) is calculated as follows:

\begin{equation}
\begin{multlined}
inStream(S_n) = \lambda^{S_n} +  \\ \mathop{\sum\nolimits_{e_x \in E | s_{dest}^x=S_n}} (\gamma^{S_{org}^x} \ast \sum\nolimits_{v \in pro(S_x)} \varphi(S_x, v) ) \ast \textyen^x \textrm{ MB/s}
\end{multlined}
\end{equation}

The following constraint of data processing should be maintained:

\begin{equation} \label{eq:3}
\sum\nolimits_{v \in pro(S_n)} \varphi(S_n, v) \geq \alpha_{S_n} 
\end{equation}
\[\textit{Where   } \alpha_{S_n} = 
\begin{cases}
\mu^{S_n}, \textit{if maximum throughput}\\

inStream(S_n), otherwise
\end{cases}
\]

\textcolor{black}{Additionally, we assume that every data stream should be processed, as unprocessed data streams lead to incorrect results. We also assume that the order of stream portions should be maintained during the distribution among corresponding compute resources.  Based on these assumptions, we maintain user specific throughputs for all services, and end-to-end latency (response time) as low as possible or even bounded when it is being increased, because if the input data rate of a service exceeded the data rate specified in processing performance constraint, the exceeded streams will be dropped. Thus, the incoming data streams upto throughput of a service are processed as they arrive, and the latency from the time of stream being added to input queue until its emission from the service as output stream is maintained. Of course, in case of a child service receives two or more dependency streams from its parents services, the latency is from the time of the last dependency stream being added to input queue until its emission from child service.}

Each service \(S_n\) in workflow application produces output stream as a result of
computation. Let \(outStream(S_n)\) denote the output data stream for a service \(S_n\) and is calculated as follows:
\begin{equation} \label{eq:5}
outStream(S_n) = \gamma^{S_n} \ast inStream(S_n) \qquad \textrm{MB/s} 
\end{equation}

\textcolor{black}{The total cost of running all provisioned VMs for all services to process incoming data streams during the period of time T seconds (which represents a set of one second interval, T = \(T_1, T_2, ..., T_I\)), is denoted as execCost(S,T) and is calculated as:}

\begin{equation}
execCost(S,T) = T \ast \sum\nolimits_{S_n} \sum\nolimits_{v \in pro(S_n)} \cent_v \qquad \textrm{cents}
\end{equation}

\textcolor{black}{The data transfer cost is based on the amount of data being moved, the cost of data transfer charged by cloud provider, and network performance. In a workflow application, both input and output data are moved among different clouds. As the speed of data may vary during workflow execution either decreases below service throughput or increases upto service throughput (as exceed load will be dropped), the calculation of data transfer cost needs to be carried-out per second. Let \(cts(S_n)\) denotes the cost of transferring streams for \(S_n\) (including input streams from other services) per second, and \(CTStream(S, T)\) denotes the total data transfer cost for the amount of data being moved for all services during the period of time T. The \(CTStream(S,T)\) is calculated as follows:}

\begin{equation}
CTStream(S, T) = \sum\nolimits_{T_i} \sum\nolimits_{S_n} cts(S_n) \qquad \textrm{cents} 
\end{equation}

\[cts(Sn) = \sum\nolimits_{S_i \in parent(S_n)} c(S_i)  \textrm{    cents}\]

\[
c(S_i) =  
\begin{cases}
0,& \text{if } C_g(S_i) = C_g(S_n)\\
outStream'(S_i) \\ \ast D(C_g(S_i),\\C_g(S_n)), & \text{otherwise}
\end{cases}
\]

\[
outStream'(S_i)
\begin{cases}
outStream(S_i),& \text{if } \varrho \leq 1\\

\frac{outStream(S_i) \ast \textyen^x}{\varrho} , & \text{otherwise}
\end{cases}
\]

\[ \textit{Where } \varrho = \frac{outStream(S_i)\ast \textyen^x}{B(C_g(S_i), C_g(S_n))} + L(C_g(S_i), C_g(S_n))\] 
\[ \textit{      , and } parent(S_n) \textit{is the set of parent services for service } S_n\]

Thus, the objective function is to minimize the cost of execution of the workflow without violating data dependences and real-time performance requirements in term of maximum throughputs: 

\begin{equation} \label{eq:8}
min f(S,T) = execCost(S,T) + ctStream(S, T)
\end{equation}

\section{Proposed Algorithms}
The problem of selecting the right resources for executing stream workflow applications in Multicloud environments to meet user requirements and to achieve efficient performance (in term of throughput and latency) while minimizing the costs of resource provisioning and data transfer is an optimization problem, where resource selection problem is generally NP-complete problem. \textcolor{black}{Our research problem is to find near-optimal resource selection solution with minimal execution cost at deployment time for executing stream workflow application in Multicloud environment, where the required resources are provisioned based on user-defined performance requirements and then services are being scheduled on these resources before the execution begins. For that, we propose two resource provisioning and scheduling algorithms using Greedy and Genetic heuristics.}

\begin{algorithm}
	\scriptsize
	\caption{Greedy Scheduling}\label{alg:GreedyAlgorithm}
	\begin{algorithmic}[1]
		\Procedure{GreedySelection}{VMOffers, unitDPRate}
		
		\For{\texttt{each service} $S_n$ \texttt{from S}}
		\State $\textit{selectedVMList} \gets \text{} \phi$
		\State $\textit{cost} \gets \text{} \infty$
		
		\State $\textit{unitMIPS} \gets \text{} unitDPRate \ast MI^{S_n} $

		\State $ reqMIPS \gets \text{get MIPS based on } \alpha_{S_n} \text{ and } unitMIPS $
		
		\For{\texttt{each cloud} $c_g$ \texttt{from C}}
		
		\If{$ S_n \textit{ is unmovable \& } c_g \neq \texttt{ placement cloud of } S_n $}
		\State $continue$
		\EndIf
		
		\State $\textit{selectedVM} \gets 0$
		\State $reqUnits = reqMIPS/unitMIPS $
		
		\State $\textit{workingVMList} \gets \text{}\textit{}\phi$
		
		\State $ VM^g\textit{} \gets \text{list of VM offers for } c_g \textit{}$

		\State $ VM^g\textit{} \gets VM^g - \{x \in VM^g \mid MIPS_x < unitMIPS\} $

		\If{$ VM^g \text{ is empty} $}
		\If{$ S_n \text{is unmovable } || S_n \text{ is movable \& } c_g \text{ is last cloud}$}
		\State{exit}
		\Else
		\State{continue}		
		\EndIf
		\EndIf

		\While{$reqUnits > 0$}
		\State $\textit{maxVMValue} \gets 0$
		\For{\texttt{each} $VM^g_k$ \texttt{from} $VM^g$}
	
		\State $\textit{achievedPortionsByVM} \gets \lfloor MIPS_{vm_k^g} / (unitMIPS) \rfloor $
	
		\State $\textit{vmValue} \gets (achievedPortionsByVM / reqUnits) / \cent_{vm_k^g} $
		
		\State $\textit{vmValue} \gets vmValue +  \lfloor MIPS_{vm_k^g} / (unitMIPS \ast \#OfServiceDependencies) \rfloor / \cent_{vm_k^g} $
		
		\If{$ vmValue > maxVMValue $}
		\State $\textit{maxVMValue} \gets vmValue $
		\State $\textit{selectedVM} \gets k$
		\EndIf
		
		\EndFor
		
		\State $\textit{workingVMList} \gets workingVMList \cup \{VM^g_{selectedVM}\} $
		\State $\textit{acheivedPortions} \gets \lfloor MIPS_{vm^g_{selectedVM}} / unitMIPS \rfloor $

		\State $\textit{reqUnits} \gets reqUnits - acheivedPortions $
		\EndWhile
		
		\State $\textit{newCost} \gets \sum\nolimits_{v \in workingVMList} \cent v $
		
		\If{$ newCost < cost $}
		\State $\textit{cost} \gets newCost v$
		\State $\textit{selectedVMList} \gets workingVMList $
		\EndIf
		
		\If{$ S_n \textit{ is unmovable \& } c_g = \textit{ placement cloud of } S_n $}
		\State $break$
		\EndIf
		
		\EndFor
		
		\State $\textit{add selectedVMList of } S_n \text{ to ServiceVMsMap}  $			
		\EndFor
		
		\EndProcedure
	\end{algorithmic}
\end{algorithm}

\subsection{Greedy Scheduling Algorithm}

A greedy algorithm is a heuristic algorithm that finds the best solution at each stage (local optimum) without consideration of future results, hoping to find global optimum. \textcolor{black}{For our resource provisioning and scheduling problem for executing stream workflow application in Multicloud environment, we propose a greedy algorithm that finds the best resource selection solution for a given workflow application at deployment time. The pseudocode of this proposed algorithm is shown in Algorithm \ref{alg:GreedyAlgorithm}.} \textcolor{black}{This algorithm takes $O(S C U V)$ with S the number of services, C the number of clouds, U the number of required minimum data processing units for a service and V the number of VM offers in the placement cloud.}

\subsection{Genetic Scheduling Algorithm}
\par \textcolor{black}{For the research problem discussed in this paper, search spaces are large and complex, with many cloud offerings available and several problem-dependent constraints to be satisfied. The search space will rapidly grow when looking for efficient schedules of increasing problem size. To deal with scheduling problem of stream workflow at deployment time, the goal is to find near-optimal solution by rapidly traversing large search spaces and generate scheduling plan for starting the execution of this workflow.}

\textcolor{black}{Genetic Algorithm (GA) is a useful algorithm to this problem because of its effectiveness at searching large and complex spaces to enable the practical implementation of optimizing scheduling. It is capable to provide several satisfying candidate solutions (i.e. resource selection solutions) to choice from by evolving over generations of candidate solutions.} Algorithm \ref{alg:GAAlgorithm} shows the pseudocode of the proposed genetic resource provisioning and scheduling algorithm. \textcolor{black}{This algorithm takes $O(G P S^2 D)$ with G the number of generations (as termination condition), P the size of population, S the length of candidate solution (number of services) and D is maximum number of stream dependencies among services. Our proposed GA} is implemented using the Watchmaker framework for evolutionary computation \cite{watchmaker2010framework}.

\textcolor{black}{The details of the proposed GA (encoding, initial population, fitness function, selection, crossover, mutation and replacement) are presented in Appendix A.}

\begin{algorithm}
	\scriptsize
	\caption{Genetic Resource Scheduling}\label{alg:GAAlgorithm}
	\begin{algorithmic}[1]

		\State $\textit{P} \gets \text{}\textit{empty initial population}$
		\State $\text{}\textit{call greedy algorithm and add its solution to P}$
		\State $\text{}\textit{generate N-1 candidates randomly and add them to P}$
		\State $\textit{calculate fitness values for candidates in P}$
		\State $\textit{sort candidates in P in ascending order of fitness}$
		
		\While{$ \text{condition not satisfied}$}
		\State $ \textit{perform elitist selection}$
		\State $ \textit{select candidates using selection operator for evolving}$
		\State $ \textit{create new offsprings using crossover operator}$		
		\State $ \textit{create new offsprings using mutation operator}$				
		\State $ \textit{replace weakest candidates using replacement opoerator}$		
		\State $ \textit{add elite candidates to the evovled population}$		
		\State $\parbox[t]{\dimexpr\linewidth-\algorithmicindent}{\textit{calculate fitness values for candidates of the evovled population}}$
		\State $\parbox[t]{\dimexpr\linewidth-\algorithmicindent}{ \textit{sort candidates of the evolved population in the ascending order of fitness}}$
		\EndWhile
		\State $ \text{return best candidate (candidate with minimum cost)}$
	\end{algorithmic}
\end{algorithm}

\section{Performance Evaluation}

\subsection{Experiment Methodology}

\subsubsection{Configuration of Workflow Application}
Common workflow structures from different application domains, such as Montage in Astronomy, Inspiral in Astrophysics, Epigenomics in Bioinformatics and Cybershake in Earthquake science, operate on static data inputs and produce outputs. The structures of these workflows and their characteristics are explained in details in \cite{bharathi2008characterization}. \textcolor{black}{Nevertheless these structures can be used as applications models to simulate different stream workflow applications after extending their XML structures. Moreover, these structures come with different sizes, so that we can conduct small to medium to large experiments with different simulated workflow structures (i.e. stream workflows)}. 

Each of these workflow structures can be used to simulate a stream workflow application, where each job is considered a service and the data flow becomes streams of data. The inputs of a job that comes from static files (not outputs of previous jobs) become the continuous inputs of a service from data producers (i.e. external sources). The service continuously processes incoming data streams and continuously produce output streams. The output of a job, which is sent to one or more jobs, becomes the continuous output of a service that is sent to one or more services. Moreover, additional parameter configurations should be added to the simulated workflow structure to make them stream workflow such as including data processing requirements, input and output data rates; these parameter configurations will be discussed in subsequent subsections. Hence, we can have workflow application with continuous inputs continuous processing and continuous outputs.

Accordingly, different stream workflow applications can be modelled using the above mentioned workflow structures for our experiments. For each workflow structure, three different sizes of such structure are used (small, medium and large) as listed in Table \ref{tab:WorkflowStructuresSizes}.

\begin{table}
	\scriptsize
	\caption{Workflow structures with their different sizes}
	\begin{tabular}{|p{3.25em}|p{6.2em}|p{5.7em}|p{6.2em}|p{5.8em}|}
		
		\hline
		Size  & Montage & Inspiral & Epigenomics & CyberShake \\
		\hline
		Small & 25 node (Montage\_25) & 30 node (Inspiral\_30) & 24 node Epigenomics\_24 & 30 node (CyberShake\_30) \\
		
		\hline
		Medium & 50 node (Montage\_50) & 50 node (Inspiral\_50) & 46 node Epigenomics\_46 & 50 node (CyberShake\_50) \\
		
		\hline
		Large & 100 node (Montage\_100) & 100 node (Inspiral\_100) & 100 node (Epigenomics\_100) & 100 node (CyberShake\_100) \\
		\hline
	\end{tabular}%
	\label{tab:WorkflowStructuresSizes}%
\end{table}%

\subsubsection{Multicloud Environment}

\textcolor{black}{We model three cloud infrastructures (Amazon EC2 \cite{Amazon2017Instances}, Google Cloud Engine \cite{Google2017Instances}, and Microsoft Azure \cite{Microsoft2017Instances}) with different VM configurations chosen from pre-defined machine types offered by those clouds. These VM configurations are provided in Table \ref{tab:VMConfigurations}. We used our simulator (IoTSim-Stream) \cite{barika2019iotsim} that is built on top of CloudSim to simulate these infrastructures as a Multicloud environment. In CloudSim \cite{calheiros2011cloudsim}, MIPS rating is used to represent CPU unit, where the capacity of a VM instance is represented by the total MIPS assigned to such instance according to the assigned value of MIPS rating multiplied by the number of assigned CPU cores (Processing Elements (PEs) in CloudSim term). Hence, the processing power of each VM instance offered by the modelled cloud is converted to the corresponding MIPS value.}

\textcolor{black}{To convert the capacity of each VM instance offered by modelled clouds to corresponding MIPS value, we use the following approach: for Amazon EC2, CPU core provides the equivalent CPU capacity of 1000 MIPS \footnote{\textcolor{black}{For Amazon EC2, one ECU provides the equivalent CPU capacity of a 1.0-1.2 GHz 2007 Opteron or 2007 Xeon processor, approximately 1000 MIPS \cite{javadi2011statistical}.}} (1 ECU), for Google Compute Engine, CPU core provides the equivalent CPU capacity with 2750 MIPS \footnote{\textcolor{black}{For Google Cloud Engine, the Google Compute Engine Unit (GCEU) is defined as a minimum processing unit, which is the equivalent one ECU \cite{ahuja2015performance}. The CPU core in Google Compute Engine provides minimum processing power equivalent to 2.75 GCEUs (2.75 ECUs), approximately 2750 MIPS \cite{ahuja2015performance}.}} (2.75 ECUs), and for Microsoft Azure, CPU core provides the equivalent CPU capacity with 2500 MIPS \footnote{\textcolor{black}{For the D1-5 v2, D2-64 v3 and F series of machine types in Microsoft Azure, these instances are based 2.4 GHz Intel Xeon® E5-2673 v3 (Haswell) processor, the 2.3 GHz Intel Xeon® E5-2673 v4 (Broadwell) processor and the 2.4 GHz Intel Xeon® E5-2673 v3 (Haswell) processor respectively \cite{Microsoft2017Instances}. Based on that, we can assume a CPU core is roughly equivalent to 2.5 ECUs, approximately 2500 MIPS.}} (2.5 ECUs).}

\begin{table}
	\tiny
	\centering
	\caption{\textcolor{black}{VM configurations of modelled clouds}}
	\begin{tabular}{|p{4em}|c|p{2em}|p{5em}|c|c|p{4em}|}
		\hline
		Cloud Provider & VM Type& \parbox{0.5cm}{vCPUs/ cores} & ECUs & Total MIPS & Memory (GB) & Price ($\cent$/second) \\
		\hline
		\multirow{11}[1]{*}{\parbox{0.8cm}{Amazon EC2 (Windows instances)}} & m4.large & 2 & 6.5 (7) & 7000 & 8 & 0.0054 \\
		& m4.xlarge & 4 & 13 & 13000 & 16 & 0.0107 \\
		& m4.2xlarge & 8 & 26 & 26000 & 32 & 0.0214\\
		& m4.4xlarge & 16 & 53.5 (54) & 54000 & 64 & \\
		& m4.10xlarge & 40 & 124.5 (125) & 125000 & 160 & 0.1067 \\
		& m4.16xlarge & 64 & 188 & 188000 & 256 & 0.1707 \\
		& c4.large & 2 & 8 & 8000 & 3.75 & 0.0054 \\
		& c4.xlarge & 4 & 16 & 16000 & 7.5 & 0.0107 \\
		& c4.2xlarge & 8 & 31 & 31000 & 15 & 0.0213 \\
		& c4.4xlarge & 16 & 62 & 62000 & 30 & 0.0426 \\
		& c4.8xlarge & 36 & 132 & 132000 & 60 & 0.0859  \\
		\hline
		\multirow{13}[1]{*}{\parbox{0.8cm}{Google Compute Engine (n1-series)}} & n1-standard-1 & 1 & 2.75 & 2750 & 3.75 & 0.0014  \\
		& n1-standard-2 & 2 & 5.5 & 5500 & 7.5 & 0.0027 \\
		& n1-standard-4 & 4 & 11 & 11000 & 15 & 0.0053  \\
		& n1-standard-8 & 8 & 22 & 22000 & 30 & 0.0106  \\
		& n1-standard-16 & 16 & 44 & 44000 & 60 & 0.0212 \\
		& n1-standard-32 & 32 & 88 & 88000 & 120 & 0.0423 \\
		& n1-standard-64 & 64 & 176 & 176000 & 240 & 0.0845 \\
		& n1-highcpu-2 & 2 & 5.5 & 5500 & 1.8 & 0.002  \\
		& n1-highcpu-4 & 4 & 11 & 11000 & 3.6 & 0.004  \\
		& n1-highcpu-8 & 8 & 22 & 22000 & 7.2 & 0.0079  \\
		& n1-highcpu-16 & 16 & 44 & 44000 & 14.4 & 0.0158 \\
		& n1-highcpu-32 & 32 & 88 & 88000 & 28.8 & 0.0316 \\
		& n1-highcpu-64 & 64 & 176 & 176000 & 57.8 & 0.0631 \\
		\hline
		\multirow{16}[1]{*}{\parbox{0.8cm}{Microsoft Azure  (Windows D and F-Series)}} & D1 v2 & 1 & 2.5 & 2500 & 3.58 & 0.0035 \\
		& D2 v2 & 2 & 5 & 5000 & 7 & 0.0069 \\
		& D3 v2 & 4 & 10 & 10000 & 14 & 0.0137 \\
		& D4 v2 & 8 & 20 & 20000 & 28 & 0.0274 \\
		& D5 v2 & 16 & 40 & 40000 & 56 & 0.052 \\
		& D2 v3 & 2 & 5 & 5000 & 8 & 0.0054 \\
		& D4 v3 & 4 & 10 & 10000 & 16 & 0.0107 \\
		& D8 v3 & 8 & 20 & 20000 & 32 & 0.0214 \\
		& D16 v3 & 16 & 40 & 40000 & 64 & 0.0427 \\
		& D32 v3 & 32 & 80 & 80000 & 128 & 0.0854 \\
		& D64 v3 & 64 & 160 & 160000 & 256 & 0.1707 \\
		& F1 & 1 & 2.5 & 2500 & 2 & 0.0027 \\
		& F2 & 2 & 5 & 5000 & 4 & 0.0054 \\
		& F4 & 4 & 10 & 10000 & 8 & 0.0107 \\
		& F8 & 8 & 20 & 20000 & 16 & 0.0213 \\
		& F16 & 16 & 40 & 40000 & 32 & 0.0426 \\
		\hline

	\end{tabular}%
	\label{tab:VMConfigurations}%
\end{table}%

\subsubsection{Network Configuration}
To model network performance (i.e. bandwidth and latency) of modelled clouds, we have conducted TCP bandwidth and latency tests between different zones of Nectar Cloud using IPerf (a cross-platform network performance measurement tool for both TCP and UDP) to collect the results for network bandwidth (in MB/s) and PING tool to collect the results for network latency (in second). From the obtained results, we create three ranges for bandwidth and latency for ingress and egress traffic as listed in Table \ref{tab:IngressBwLat} and \ref{tab:EgressBWLat} respectively.

\begin{table}[H]
	\scriptsize
	\centering
	\caption{Ranges of ingress network bandwidth and latency.}

	\begin{tabular}{|p{5em}|p{13em}|p{13em}|}

		\hline

		Range & Minimum Bandwidth (MB/s) / Latency (seconds) & Maximum Bandwidth (MB/s) / Latency (seconds) \\

		\hline
		Low   & 302 / 0.0004 & 614 / 0.00063 \\
		
		\hline
		Medium & 615 / 0.00064 & 926 / 0.00086 \\
		
		\hline
		High  & 927 / 0.00087 & 1238 / 0.0011 \\
		
		\hline
	\end{tabular}%
	\label{tab:IngressBwLat}%
\end{table}%

\begin{table}[H]
	\scriptsize
	\centering
	\caption{Ranges of egress network bandwidth and latency.}

	\begin{tabular}{|p{5em}|p{13.0em}|p{13em}|}

		\hline

		Range & Minimum Bandwidth (MB/s) / Latency (seconds) & Maximum Bandwidth (MB/s) / Latency (seconds) \\
		
		\hline
		Low   & 24 / 0.009 & 121 / 0.020 \\
		
		\hline
		Medium & 122 / 0.021 & 218 / 0.031 \\
		
		\hline
		High  & 219 / 0.032 & 314 / 0.040 \\
		
		\hline
	\end{tabular}%
	\label{tab:EgressBWLat}%
\end{table}%

\subsubsection{Data Transfer Cost}
For Internet egress traffic, the cost/rate of data transfer for each modelled cloud is based on the monthly usage tier and the destination zone. To model the costs of data transfer (in cents/MB) for our experiments, we find the minimum and maximum data transfer costs between modelled clouds, and then use them to create three ranges (low, medium and high) as listed in Table \ref{tab:DTCOST}. For ingress traffic, the cost is 0. 

\begin{table}[H]
	\scriptsize
	\centering
	\caption{Ranges of outbound data transfer cost for clouds}
	\begin{tabular}{|p{10.0em}|p{10.0em}|p{10.5em}|}
		
		\hline
		
		Range & \multicolumn{1}{c|}{Minimum (cents/MB)} & \multicolumn{1}{c|}{Maximum (cents/MB)} \\
		
		\hline
		Low     & 0.005   & 0.012 \\
		
		\hline
		Medium  & 0.013   & 0.019 \\
		
		\hline
		High    & 0.020    & 0.025 \\
		
		\hline
	\end{tabular}%
	\label{tab:DTCOST}%
\end{table}%

\subsubsection{Data Rate of External Source}
To model data rate of external sources (IoT devices such as sensor), we choose minimum and maximum data rate based on different data rates of various technologies/standards of IoT defined in \cite{IoTStandardsandProtocols}, where the minimum is 0.0013 MB/s and maximum is 12.5 MB/s. From the chosen minimum and maximum, we create three different data rate ranges for our experiment, as listed in Table \ref{tab:EXSourceDataRate}.

\begin{table}[H]
	\scriptsize
	\centering
	\caption{Ranges of external source data rate}
	\begin{tabular}{|p{10.0em}|p{10.0em}|p{10.5em}|}
		
		\hline
		
		Range & Minimum (MB/s) & Maximum (MB/s) \\
		
		\hline
		Low     & 0.0013 (≈10.7 Kbps) & 4.2 (33.6 Mbps) \\
		
		\hline
		Medium  & 4.3 (34.4 Mbps) & 8.4 (67.2 Mbps) \\
		
		\hline
		High    & 8.5 (68Mbps) & 12.5 (100Mbps) \\
		
		\hline
	\end{tabular}%
	\label{tab:EXSourceDataRate}%
\end{table}%

\subsubsection{Types of Service}
Since each service of workflow application can be either movable or unmovable, there is a need to determine how many of those services are movable and how many of those services are unmovable. For unmovable services, we need to specify the placement cloud for each one of them. By considering workflow application as strict application, the number of movable services are low compared with the number of unmovable services. With more flexible and pervasive workflows, the number of movable services are high compared with low number of unmovable services, where there is a possibility for this type of workflow application to be all of its services are movable. Thus, by considering the different natures of strict workflow applications and highly flexible and pervasive workflow applications, we create three percentages ranges of movable services in workflow application and listed them in Table \ref{tab:MovableServices}.

\begin{table}[H]
	\scriptsize
	\centering
	\caption{Percentage ranges of movable services}
	\begin{tabular}{|p{10.0em}|p{10.0em}|p{10.5em}|}
		
		\hline
		
		Range & Minimum (\%) & Maximum (\%) \\
		
		\hline
		
		Low     & 0\% & 34\% \\
		
		\hline
		
		Medium  & 35\% & 68\% \\
		
		\hline

		High    & 69\% & 100\% \\

		\hline
	\end{tabular}%
	\label{tab:MovableServices}%
\end{table}%

\subsubsection{Data Processing Requirement of Services}
To model data processing requirement for services (simple or/and complex services), we create different ranges for data processing requirement as listed in Table \ref{tab:ServiceDPReq}, based on the following specified minimum and maximum values: the minimum value for data processing requirement for a service is 20 MI/MB (representing data processing requirement for simple aggregation functions) and the maximum value for data processing requirement for a service is 4000 MI/MB (representing data processing requirement for complex aggregation functions).

\begin{table}[H]
	\scriptsize
	\centering
	\caption{Ranges of service data processing requirement}
	\begin{tabular}{|p{10.0em}|p{10.0em}|p{10.5em}|}

		\hline
		
		Range & \multicolumn{1}{p{10.0em}|}{Minimum (MI/MB)} & \multicolumn{1}{p{10.5em}|}{Maximum (MI/MB)} \\
		
		\hline
		Low     & 20      & 1347 \\
		
		\hline
		Medium  & 1348    & 2674 \\
		
		\hline
		High    & 2675    & 4000 \\
		
		\hline
	\end{tabular}%
	\label{tab:ServiceDPReq}%
\end{table}%

\subsubsection{Output Data Rate of Service}
As the output data rate of a service is calculated using Equation \ref{eq:5}, specification of the proportion of data generated by a service based on input streams can be used to model output data rates for services in workflow applications. For modelling different ranges of service output data rate, we define the minimum and maximum output proportion/percentage generated by service based on input streams to be 0.01/1\% and 1.5/150\% respectively, and use them to create three ranges for service output data rate as listed in Table \ref{tab:ServiceOutput}.

\begin{table}[H]
	\scriptsize
	\centering
	\caption{Percentage ranges of service output data rate}

	\begin{tabular}{|p{5.0em}|p{13.0em}|p{13em}|}

		\hline

		Range & Minimum (proportion / \%) & Maximum (proportion / \%) \\

		\hline
		Low     & 0.01 / 1\% & 0.50 / 50\% \\

		\hline
		Medium  & 0.51 / 51\% & 1.0 / 100\% \\

		\hline
		High    & 1.01 / 101\% & 1.5 / 150\% \\

		\hline
	\end{tabular}%
	\label{tab:ServiceOutput}%
\end{table}%

\subsubsection{Data Processing Rate for Minimum Stream Unit}
In workflow applications, the data rates streaming from different sources (either external sources or other services) as inputs to service are varied. Thus, to process these streams using compute resources of such service, these streams should be divided into portions and then be scheduled on those resources for processing. To achieve that, we need to determine the smallest stream unit per second that will be processed by each provisioned compute resource, where compute resource can process multiple units per second according to its computing capacity per second (MIPS). By specifying minimum stream unit for the whole workflow application, the data processing rate (MB/s) for processing this unit can be determined to ensure that each provisioned compute resource at least processes one unit per second. For our experiment, we create three ranges for data processing rate of minimum stream unit as listed in Table \ref{tab:MinDPRate}.

\begin{table}[H]
	\scriptsize
	\centering
	\caption{Ranges of data processing rate of minimum unit}
	\begin{tabular}{|p{13.0em}|p{8.5em}|p{8.5em}|}
		
		\hline
		
		Range & \multicolumn{1}{p{8.5em}|}{Minimum (MB/s)} & \multicolumn{1}{p{8.5em}|}{Maximum (MB/s)} \\

		\hline
		Low     & \multicolumn{1}{p{8.5em}|}{0.2 (=1.6Mbps)} & 1.0 \\

		\hline
		Medium  & 1.1     & 2.0 \\

		\hline
		High    & 2.1     & 2.9 \\

		\hline
	\end{tabular}%
	\label{tab:MinDPRate}%
\end{table}%

\subsubsection{Genetic Algorithm Configuration}
To produce results in GA, we configure its parameters as follows: population size and generation limit are 50, elitism is 1, and the probability for crossover, mutation and replacement operations are 0.8, 0.3 and 0.2 respectively.

\subsubsection{Other Simulation Parameters}
The other parameters including data processing rate ($\alpha_{S_n}$) and incoming data mode towards a service from its parent service(s) as inputs are fixed for all scenarios, and their values are system-calculated rate and replica respectively. The simulation time for all experiments is 3 minutes (180s).

\subsubsection{Experiments and Scenarios}

\textcolor{black}{To evaluate the efficiency of the proposed algorithms (Greedy and GA) in term of execution costs, and study their behaviours in term of computational time and end-to-end latency, two sets of experiments are conducted}

\textcolor{black}{First set of experiments (execution cost comparison with lower bound and fair-share method): We compare the results of execution costs obtained from the proposed algorithms (Greedy and GA) for executing the 12 modelled workflow applications with lower bound under varying of seven parameters. These parameters are data rate of external source (P1), data processing requirement of service (P2), output data rate of service (P3), type of service (P4), network bandwidth and latency (P5), cost of data transfer (P6) and data processing rate of minimum stream unit (P7). Thus, seven experimental scenarios are considered in this evaluation as shown in Table \ref{tab:Scenarios}, where in each scenario, the low, medium and high ranges of the variable parameter will be studied. In regards to lower bound, we have relaxed many constraints including service’s datacenter placement constraint, VM provisioning constraint (selecting the cheapest VM across all datacenter VM offers), data transfer cost (using a lower cost value from the studied range) and network bandwidth constraint (using a lower bandwidth from the studied range which leads to reduction in data transfer cost by transferring less data). Then for each service, the cheapest VM from the placement cloud of this service is provisioned as many as is required to achieve the specified data processing rate. After that, the total execution cost (provisioning cost + data transfer cost) is calculated using Equation \ref{eq:8} during the period of time T.} \textcolor{black}{In addition, we compare the proposed algorithms with default scheduling method used in Apache YARN and Mesos. Apache YARN uses default Fair scheduling method to equal share cluster resources between applications over time. Apache Mesos is a cluster manager, where the default scheduling decision used by the master process to determine how resources will be assigned to each framework is Dominant Resource Fairness algorithm; this algorithm is a fair sharing model to multiple resource types. Therefore, we have implemented fair-share scheduler (which provisions the same VM as many as is required to achieve the specified data processing rate for all services in a workflow). Then, we compare the results produced by this scheduler with the results from the proposed algorithms. In the aforementioned comparisons, we consider the results obtained from lower bound as the base values.}

\textcolor{black}{Second set of experiments (proposed algorithms comparison): We use the computational time and average end-to-end latency recorded from the aforementioned scenarios to study and compare behaviours of the proposed algorithms for executing different stream workflow applications.}

\begin{table}
	\scriptsize
	\centering
	\caption{Scenarios of our experimental study}
	\begin{tabular}{|l|l|l|}
		
		\hline
		\multicolumn{1}{|p{6.0em}|}{Scenario} & \multicolumn{1}{p{14em}|}{Fixed Parameters*} & \multicolumn{1}{p{9.0em}|}{Variable Parameter} \\
		
		\hline
		{Scenario 1} & {P2$-$P7} & {P1} \\
		
		\hline
		{Scenario 2} & {P1 and P3$-$P7} & {P2} \\
		
		\hline
		{Scenario 3} & {P1$-$P2 and P4$-$P7} & {P3} \\
		
		\hline
		{Scenario 4} & {P1$-$P3 and P5$-$P7} & {P4} \\
		
		\hline
		{Scenario 5} & {P1$-$P4 and P6$-$P7} & {P5} \\
		
		\hline
		{Scenario 6} & {P1$-$P5 and P7} & {P6} \\
		
		\hline
		{Scenario 7} & {P1$-$P6} & {P7} \\
		
		\hline
		\multicolumn{3}{|l|}{*The values of fixed parameters are obtained from their medium ranges} \\
		
		\hline
	\end{tabular}%
	\label{tab:Scenarios}%
\end{table}%

\subsection{Experimental Results and Discussion}
\par For our experiments, we designed and implemented IoTSim-Stream, our extended version of CloudSim \cite{calheiros2011cloudsim} that enables the execution of stream workflow applications in Multicloud environments. \textcolor{black}{The experimental scenarios are simulated to evaluate and compare the proposed algorithms with lower bound as well as with each others. In regard to the experimental results of average end-to-end latency, these results are collected after the system warmed-up (i.e. at second 120) to study the delay when simulation system is under highest pressure}. For GA, we run each experimental scenario ten times, and average results are obtained and used in representation of experimental results. 

\textcolor{black}{For space reasons, we have examined the results of all scenarios looking for those results that have little difference or have similar behaviour, and those with different behaviours. In regards to experimental results for execution cost comparison, we found that the results of Scenario 1 and 2 can be represented by the result of Scenario 2 as it expresses the highest values. Similarly, the results of Scenario 5 and 6 can be represented by the result of Scenario 6 as it expresses the highest vales. Therefore, the execution cost results of Scenario 2, 3, 4, 6 and 7 will be presented. Moreover and for the algorithm comparison using average end-toend latency, we found that the end-to-end latency results of Scenario 1 \& 2 \& 5 \& 6 have somewhat close behaviour with slight difference, therefore the result of Scenario 2 can be used to represent their behaviours as it represents the highest values. Therefore, the average latency results of Scenario 2, 3, 4 and 7 will be presented. } 

\begin{figure*}
	\centering
	\includegraphics[width=1\linewidth]{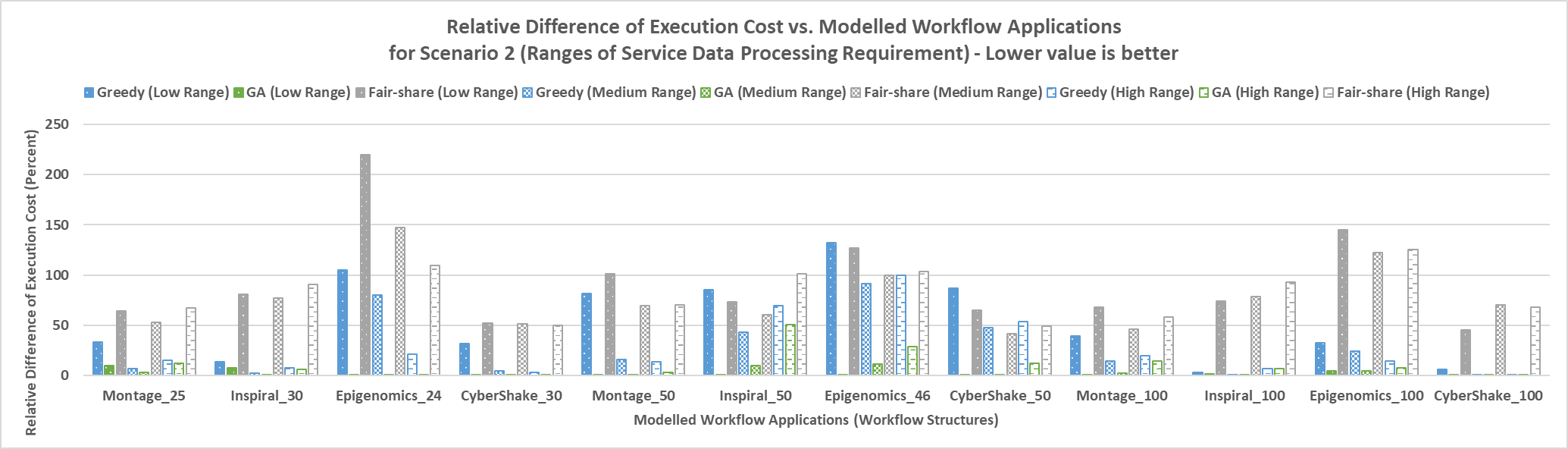}
	\caption{Execution Cost Comparison for Scenario 2.}
	\label{fig:scenario2cost}
\end{figure*}

\begin{figure*}
	\centering
	\includegraphics[width=1\linewidth]{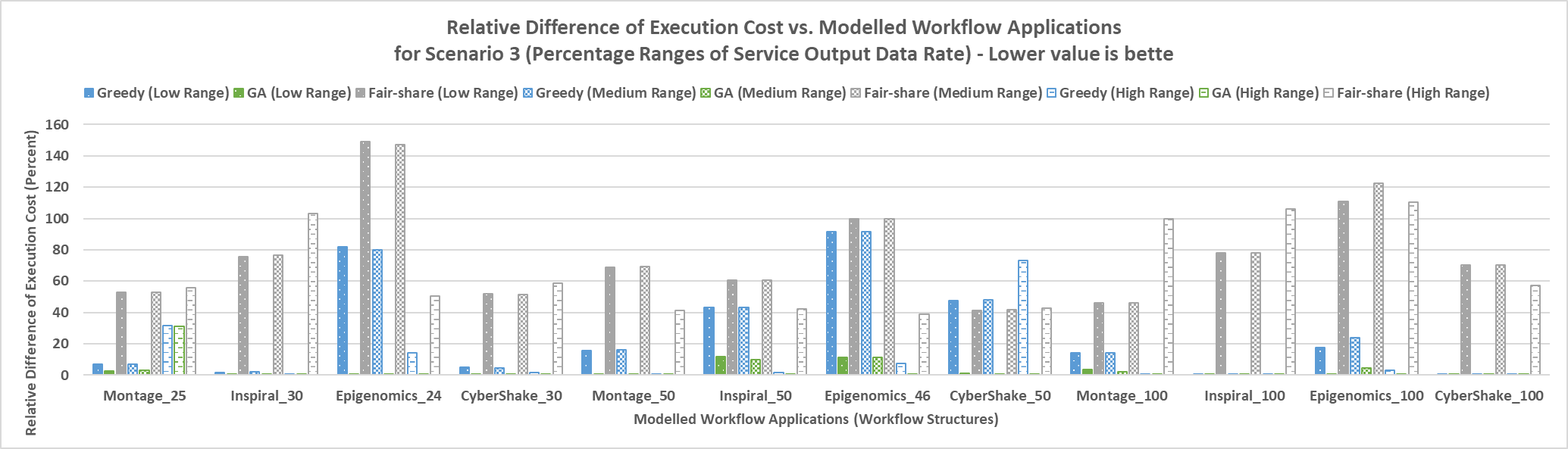}
	\caption{Execution Cost Comparison for Scenario 3.}
	\label{fig:scenario3cost}
\end{figure*}

\begin{figure*}
	\centering
	\includegraphics[width=1\linewidth]{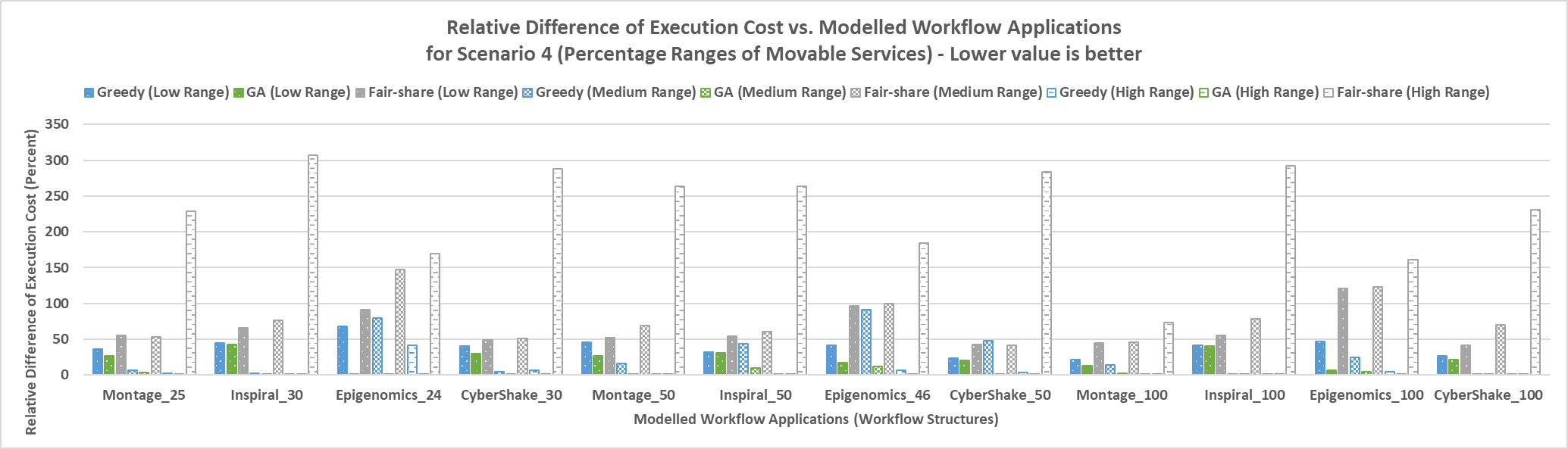}
	\caption{Execution Cost Comparison for Scenario 4.}
	\label{fig:scenario4cost}
\end{figure*}
\begin{figure*}
	\centering
	\includegraphics[width=1\linewidth]{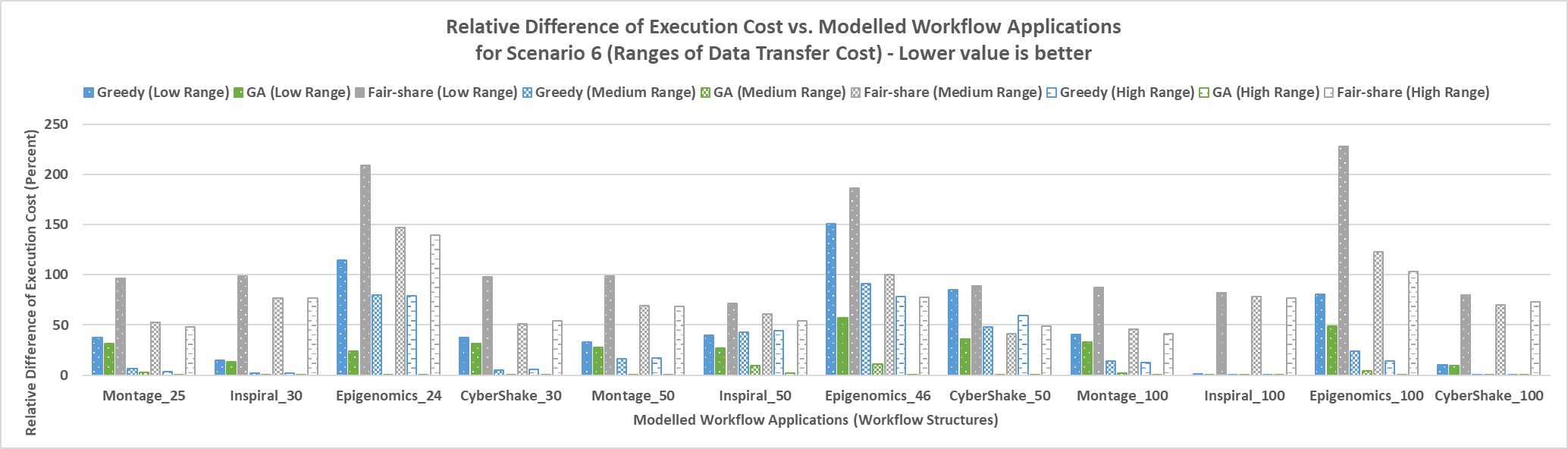}
	\caption{Execution Cost Comparison for Scenario 6.}
	\label{fig:scenario6cost}
\end{figure*}
\begin{figure*}
	\centering
	\includegraphics[width=1\linewidth]{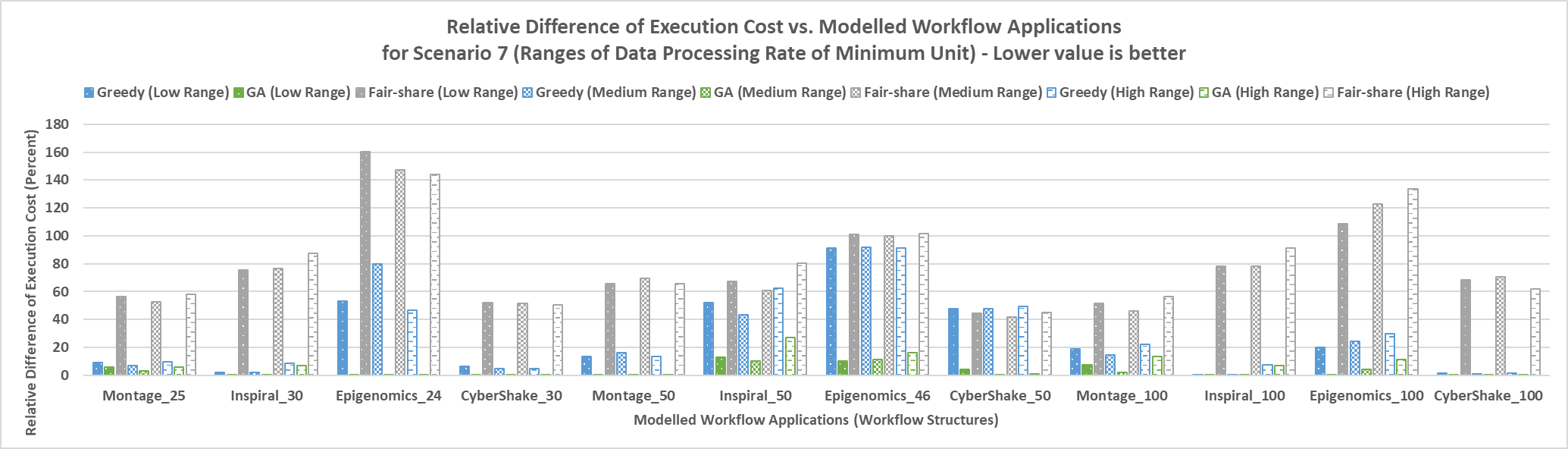}
	\caption{Execution Cost Comparison for Scenario 7.}
	\label{fig:scenario7cost}
\end{figure*}

\begin{figure*}
	\centering
	\includegraphics[width=1\linewidth]{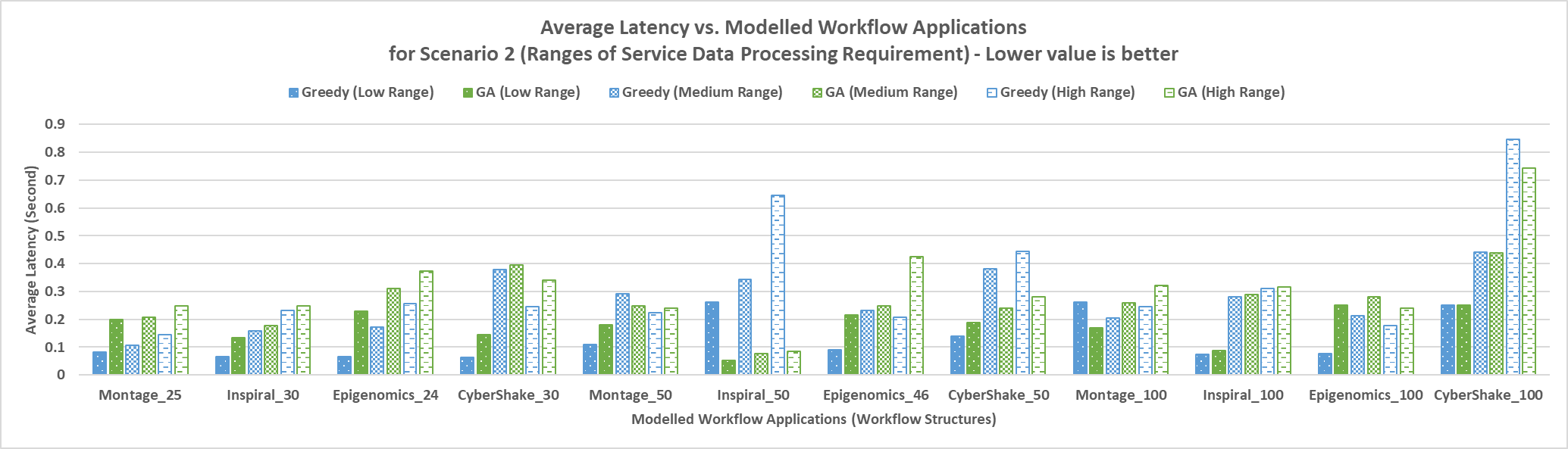}
	\caption{Proposed algorithms comparison using average end-to-end latency for Scenario 2.}
	\label{fig:scneario2latency}
\end{figure*}

\begin{figure*}
	\centering
	\includegraphics[width=1\linewidth]{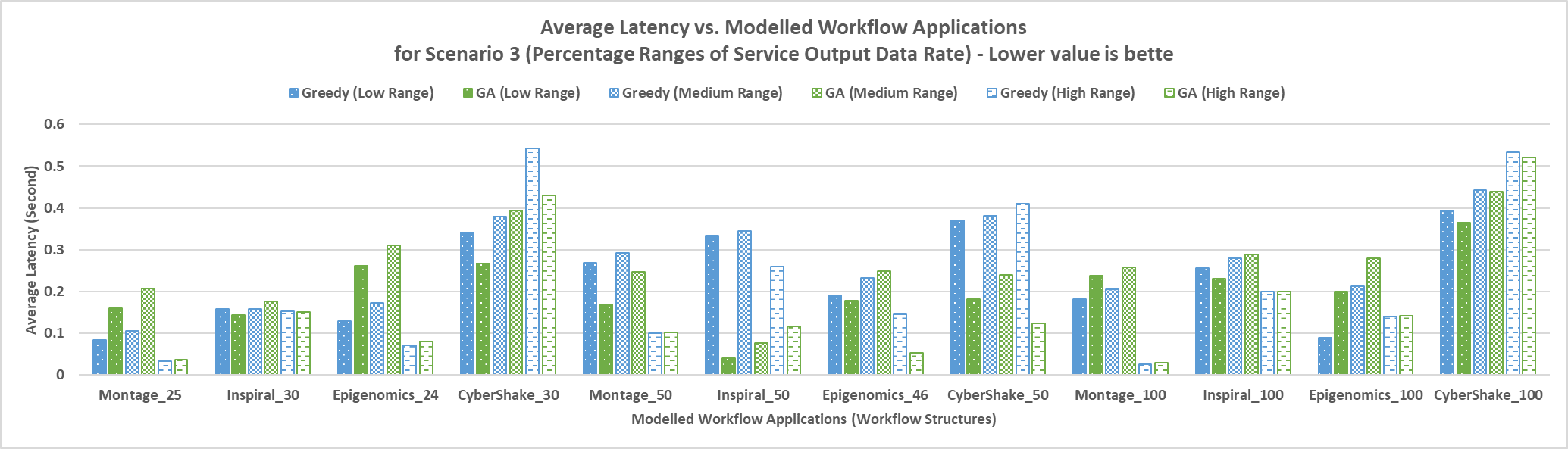}
	\caption{Proposed algorithms comparison using average end-to-end latency for Scenario 3.}
	\label{fig:scneario3latency}
\end{figure*}

\begin{figure*}
	\centering
	\includegraphics[width=1\linewidth]{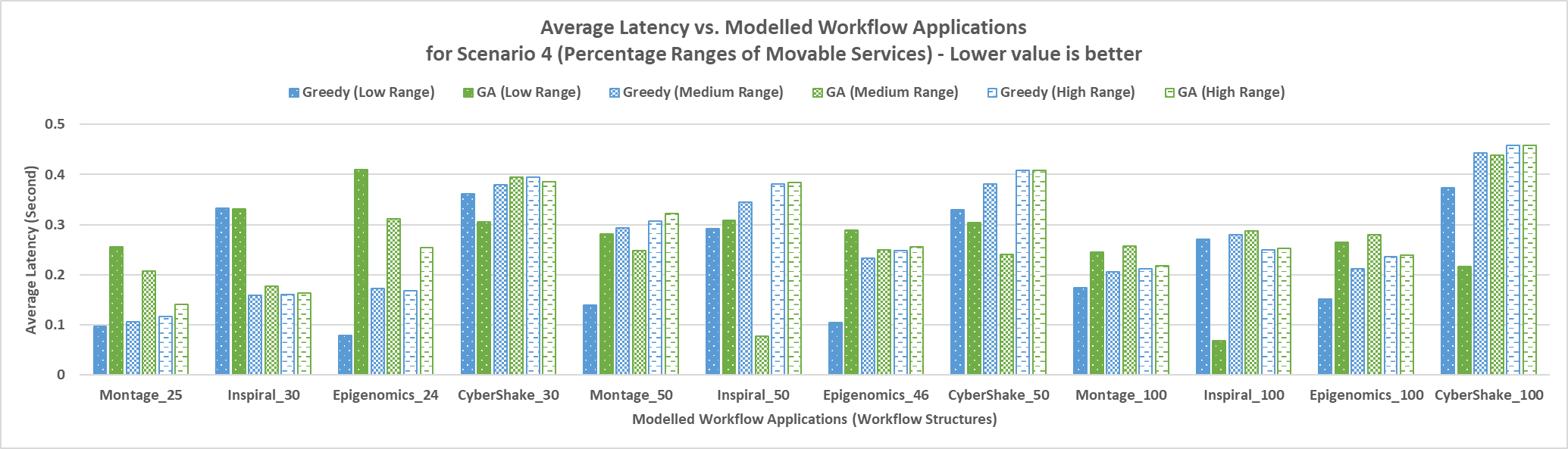}
	\caption{Proposed algorithms comparison using average end-to-end latency for Scenario 4.}
	\label{fig:scneario4latency}
\end{figure*}

\begin{figure*}
	\centering
	\includegraphics[width=1\linewidth]{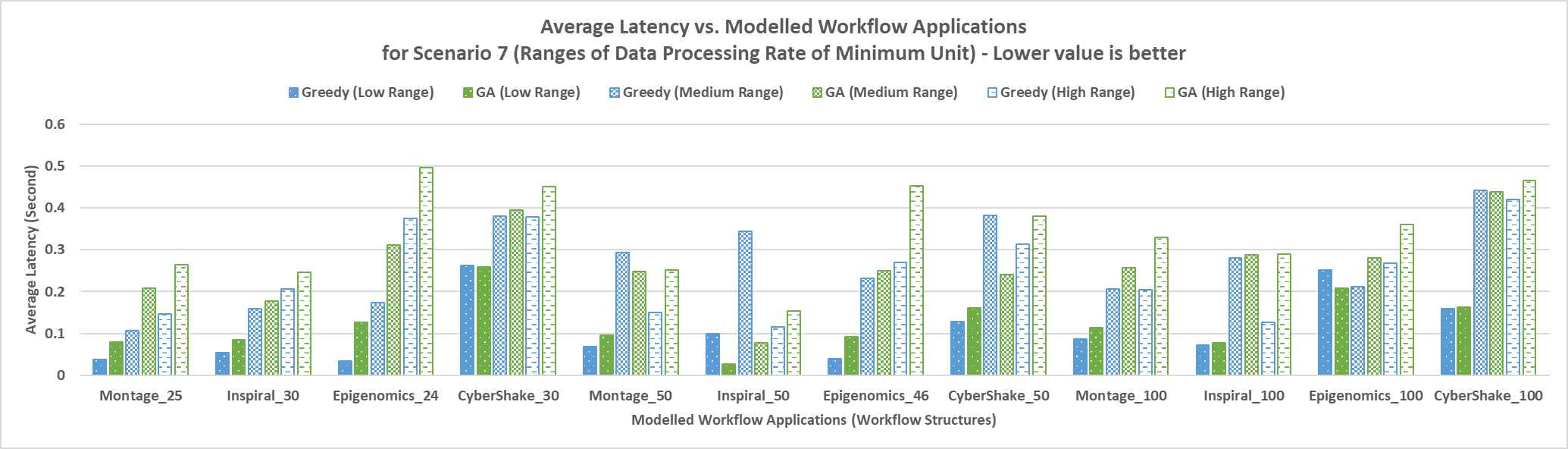}
	\caption{Proposed algorithms comparison using average end-to-end latency for Scenario 7.}
	\label{fig:scneario7latency}
\end{figure*}

\textcolor{black}{Figure \ref{fig:scenario2cost} to Figure \ref{fig:scenario7cost} depict the experimental results for the relative difference (in percentage) that achieved by the proposed algorithms in comparison to lower bound in term of execution cost. From the experimental results shown in these figures, our analysis and findings are summarized into three discussion points (DPs).}

\paragraph*{DP1}\textcolor{black}{As we expected, the presented results in these figures showed that the proposed GA achieved lowest relative differences of execution cost in comparison to greedy algorithm and fair-share method. This is clear due to GA being efficient at searching large and complex spaces by rapidly traversing these spaces and finding several satisfying candidate solutions (i.e. resource selection solutions) to choose from by evolving over generations of candidate solutions. GA surpasses the greedy algorithm in term of cost reduction by finding the best resource provisioning and scheduling solution with minimal execution cost (from those satisfying solutions) for different modelled workflow applications.} Moreover, the relative differences of execution cost obtained by the proposed GA are low in most cases, which makes this algorithm produces total execution cost results that are close to the results of the most relaxed lower bound. Of course, in some cases, there is still a difference because of the lower bound produced unachievable results. The reason for that is the proposed GA considers both costs of resource provisioning and data transfer for each candidate solution that being generated in comparison to greedy algorithm which finds a solution that reduces only resource provisioning cost and ignoring the contribution of data transfer cost and then based on that solution, the data transfer cost is calculated and added to provisioning cost making the execution cost.
	
\paragraph*{DP2}\textcolor{black}{In very few cases (such as high range in Scenario 2 with Inspiral\_100 and low range in Scenario 4 with Inspiral\_50) where the relative difference of execution cost between the proposed greedy and GA is slight, this little cost reduction is still reasonable and can be considered as an extra cost-saving when workflow application runs for several minutes, hours or even longer. For instance, high range in Scenario 2 with Inspiral\_100, the  cost-saving of running this application for just a hour is $\approx$ (\$10.44).}

\paragraph*{DP3}\textcolor{black}{By observing Figures \ref{fig:scenario4cost} and \ref{fig:scenario6cost}, in some cases with low range, the relative difference of execution cost achieved by the proposed GA is not so close to unachievable lower bound. In relation to Scenario 4, the low percentage range of movable services means that there are high placement restrictions as most of services in workflow applications are unmovable, so that the opportunity of cost reduction is narrow and mainly based on the small number of movable services, leading to GA may not be able to find near-optimal solution for executing given workflow application. Whereas with high range, GA has an ample opportunity to find near-optimal provisioning and scheduling solution that leading to total execution cost results are closer to lower bound. In relation to Scenario 6, the reason behind that is when the cost of transferring data is low, GA may face a local optimality problem since changing the provisioning plan will not adjust the contribution of data transfer cost to the total cost as it is very low in origin.}

\textcolor{black}{For proposed algorithms comparison, the computational time results expressed the straightforward conclusion, which is the greedy algorithm takes less time to generate a scheduling plan compared with genetic algorithm, but we found that genetic algorithm needs relatively low time to compute and find such plan (at most across all scenarios). Therefore, we do not need to present these results and we only present the minimum and maximum computational time (in milliseconds) for each proposed algorithm with each scenario (see Table \ref{tab:ComputationTimeResults}). In relation to average end-to-end latency, Figure \ref{fig:scneario2latency} to Figure \ref{fig:scneario7latency} show the average latency results achieved by these algorithms. Our analysis from these figures are summarized into three DPs:}

\paragraph*{DP4}\textcolor{black}{It is clear that both algorithms are able to achieve sub-second average latency for 12 modelled workflow applications with all scenarios. The proposed greedy algorithm in most cases achieved lower average latency compared with GA. The reason behind that greedy algorithm is more oriented to provision each VM that not only achieve processing the minimum stream unit based on service data processing requirement but also has compute power to process the number of minimum units that being received as input stream portions to the service. However, as mentioned earlier, GA maintains sub-second average latency across all scenarios. It even achieved lower average latency in some cases compared to greedy algorithm such as in Scenario 3 with Montage\_50, Inspiral\_50 and CyberShake\_50.} 

\paragraph*{\textcolor{black}{DP5}}\textcolor{black}{In most cases, the end-to-end latency of GA is lower than that of Greedy algorithm such as Inspiral\_50 in Figure 12 and 14, and Cybershark\_50 in Figure 12 and 13. The reason behind that is the GA is designed to utilize data locality for all services within stream workflow. This minimizes end-to-end latency by reducing data movement across multiple clouds and trying to avoid data transfer cost and time. For some cases, it is not applicable to achieve data locality due to several constrains such as huge number of data sources and their fixed placements.}

\paragraph*{DP6}\textcolor{black}{The proposed algorithms are able to achieve the maximum throughputs that defined by the owner of workflow without affecting end-to-end latency and keeping average latency in sub-second since every data stream arrives is processed as soon as the dependency is achieved. The variations in the measured average latency occur because of the structure of workflow and the data dependency relations among services that are presented in this structure.} 
	
\textcolor{black}{From the overall discussion in both comparisons, we found that GA achieved the best execution cost reduction, inexpensive computational time and good average latency while greedy algorithm achieved expensive execution cost, very low computation time and low average latency. For real-time data processing applications, end users are mainly concerned about the latency, but the expensive execution cost for the application is believed to be a barrier because of this application processes big data that need also large computational power. By considering the trade-off between the benefits of reduction in total execution cost and maintaining low computational time and end-to-end latency, we think that it is reasonable and practical to have low execution cost with little defer in average latency (bounded by a second) and computational time (bounded by several seconds). Thus, we can claim that GA is the best choice for meeting user performance requirements at deployment time while maintaining efficient performance (maximum throughput and sub-second latency) with minimal execution cost.}

\begin{table}
	\scriptsize
	\centering
	\caption{Computational Time Results (in Milliseconds)}
	\begin{tabular}{|p{7em}|p{4em}|p{4em}|p{4em}|p{4em}|}
		\hline
		& \multicolumn{2}{c|}{Greedy} & \multicolumn{2}{c|}{GA} \\
		& Min & Max & Min & Max \\
		\hline
		Scenario 1 & 1 & 219 & 57.6 & 2383.9 \\
		\hline
		Scenario 2 & 1 & 391 & 65.8 & 2910.8 \\
		\hline
		Scenario 3 & 1 & 2453 & 64.1 & 20127.9 \\
		\hline
		Scenario 4 & 1 & 219 & 65.8 & 2383.9 \\
		\hline
		Scenario 5 & 1 & 219 & 65.8 & 2653.9 \\
		\hline
		Scenario 6 & 1 & 219 & 65.8 & 3280.4 \\
		\hline
		Scenario 7 & 1 & 219 & 65.5 & 2383.9 \\
		\hline
		Median & 1 & 219 & 65.8 & 2653.9 \\
		\hline
	\end{tabular}%
	\label{tab:ComputationTimeResults}%
\end{table}%

\section{Conclusion}
\par In this paper, we modelled stream workflow applications and proposed two resource provisioning and scheduling algorithms (greedy and genetic) for efficient execution of such workflows in Multicloud environments. We also simulated different stream workflows using common workflow structures to examine the efficiency of the proposed algorithms in simulation environment using IoTSim-Stream. The experimental results obtained from our experiments showed that the proposed algorithms reduce the execution cost with modelled workflow applications, maintain throughputs and achieve sub-second latency, where the proposed GA is outperformed greedy algorithm for all experiment scenarios.

\par Our work reveals new two directions for future study. The first direction is supporting the execution of dynamic stream workflow application in Multicloud environment by either improving these techniques or proposing a new technique to dynamically adapt the scheduling and provisioning plan at runtime. The second direction is improving the proposed greedy to achieve high cost reduction, and GA with advanced operators such as global and local competition operators to further enhance its efficiency.

\vspace{-0.13in}

  \section*{Acknowledgment}
The authors would like to thank Prof. Rajkumar Buyya for the insightful comments and suggestions in improving paper quality. This research is supported by an Australian Government Research Training Program (RTP) Scholarship.

\bibliographystyle{IEEEtran}
\bibliography{references}

\appendices

\section*{Appendix}

\subsection{Details of Proposed Genetic Scheduling Algorithm}

\begin{figure*}[t!]
	\begin{subfigure}{1\textwidth}
		\centering
		\includegraphics[width=0.5\linewidth]{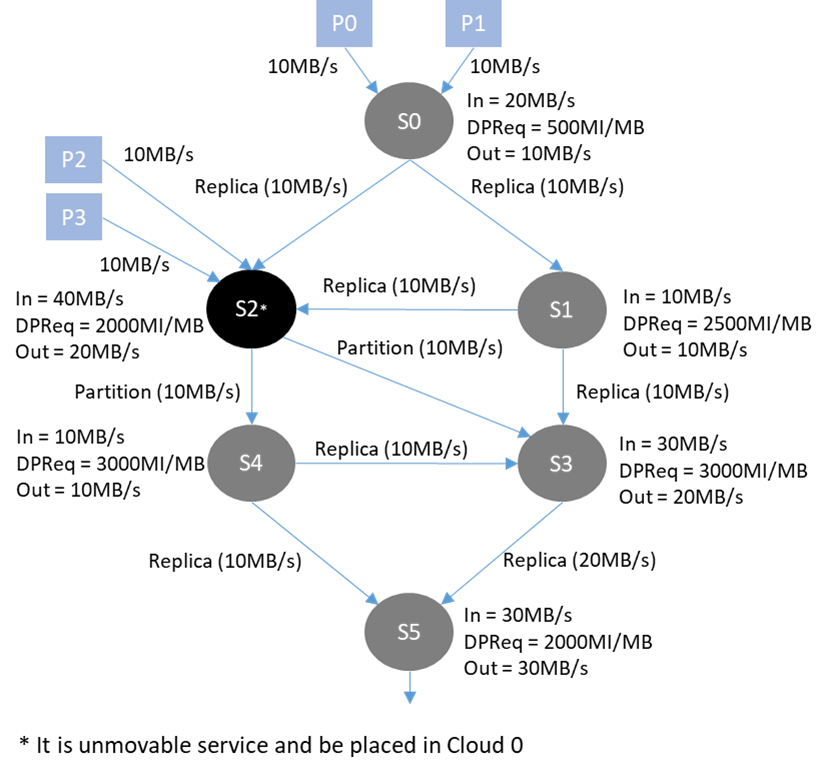}
		\caption{Sample stream workflow application}
		\label{fig:exampleworkflow}
	\end{subfigure}
	
	\hspace{0.1cm}
	\begin{subfigure}{0.5\textwidth}
		\centering
		\includegraphics[width=0.6\linewidth]{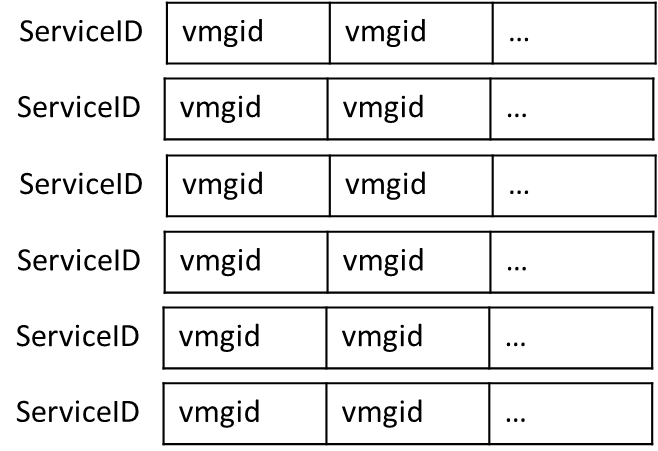}
		\caption{General candidate encoding}
		\label{fig:candidateRepresentation}
	\end{subfigure}
	\hspace{0.1cm}
	\begin{subfigure}{0.5\textwidth}
		\centering
		\includegraphics[width=0.6\linewidth]{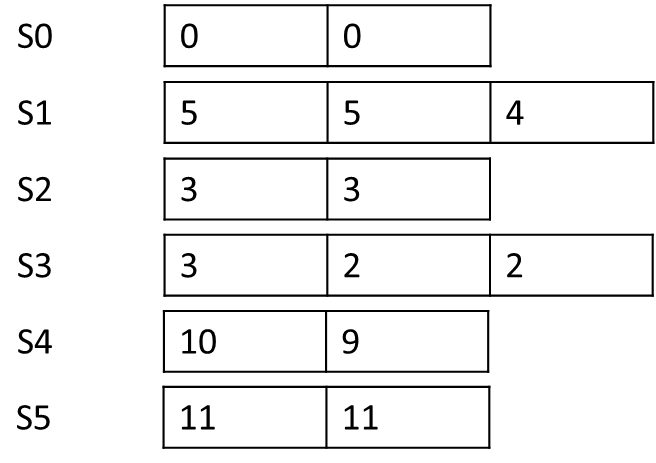}
		\caption{Sample candidate encoding}
		\label{fig:possibleCandidate}
	\end{subfigure}
	\caption{Candidate solution encoding for sample stream workflow application.}
\end{figure*}

\subsubsection{Encoding} A feasible resource selection solution composes of a set of chromosomes, where each chromosome is a data structure in which a resource selection for a service is encoded. 
The encoding of this candidate solution for the presented sample stream workflow application (Figure \ref{fig:exampleworkflow}) is depicted in Figure \ref{fig:candidateRepresentation}. We assume that the selected VMs for a service in a candidate solution should be from one cloud, where different instances of VMs can be selected as well as the same VM can be reselected many times (reputation is allowed). To deal with multiple clouds, the identifiers of the offered VMs from these clouds should be globally unique, thus they can be used conveniently in genes of chromosome's without any possible conflict. Therefore, we create a global VM mapping that map each VM offered by each cloud to global VM identifier. For instance, Table \ref{tab:globalVMMapping} shows the global VM mapping for VMs offered by three clouds. 
Based on this mapping, the two-dimensional integer encoding of possible candidate solution for sample workflow application is shown in Figure \ref{fig:possibleCandidate}. Using this encoding makes the genetic manipulations easier, for example to apply crossover, we simply swap chromosomes (services) between two candidates according to crossover points without any iteration over chromosome's genes.

\subsubsection{Initial Population}

It contains greedy solution as a one chromosome and N-1 chromosomes that are randomly generated, making the search space covering a wide range of possible resource selection solutions.

\begin{table}[H]
	\scriptsize
	\centering
	\begin{center}
		\caption{The global VM mapping for clouds}
		\begin{tabular}{|p{1.5cm}|p{2cm}|p{1.5cm}|p{2cm}|}
			\hline 
			vmgid & vmidAtCloud & CloudID & TotalMIPS \\ 
			\hline 
			0 & 0 & 0 & 7000  \\ 
			\hline 
			1& 1 & 0 & 13000 \\ 
			\hline 
			2& 2 & 0 & 26000 \\ 
			\hline 
			3& 3 & 0 & 54000 \\ 
			\hline 
			4& 0 & 1 & 5500 \\ 
			\hline 
			5& 1 & 1 & 11000 \\ 
			\hline 
			6& 2 & 1 & 22000\\ 
			\hline 
			7& 3 & 1 & 44000\\ 
			\hline 
			8& 0 & 2 & 5000\\ 
			\hline 
			9& 1 & 2 & 10000\\ 
			\hline 
			10& 2 & 2 & 20000\\ 
			\hline 
			11& 3 & 2 & 40000\\ 
			\hline 
		\end{tabular}
		\label{tab:globalVMMapping}
	\end{center}
\end{table}

\subsubsection{Fitness Function}
The fitness value for a candidate solution is computed using Equation 7.

\subsubsection{Selection}
Before making any selection in each generation, elitist selection is performed, then the simplest roulette-wheel selection is used to select candidate solutions for the next generation.

\begin{figure*}
	\begin{minipage}[t]{1\linewidth}
		\centering
		\includegraphics[width=0.7\linewidth]{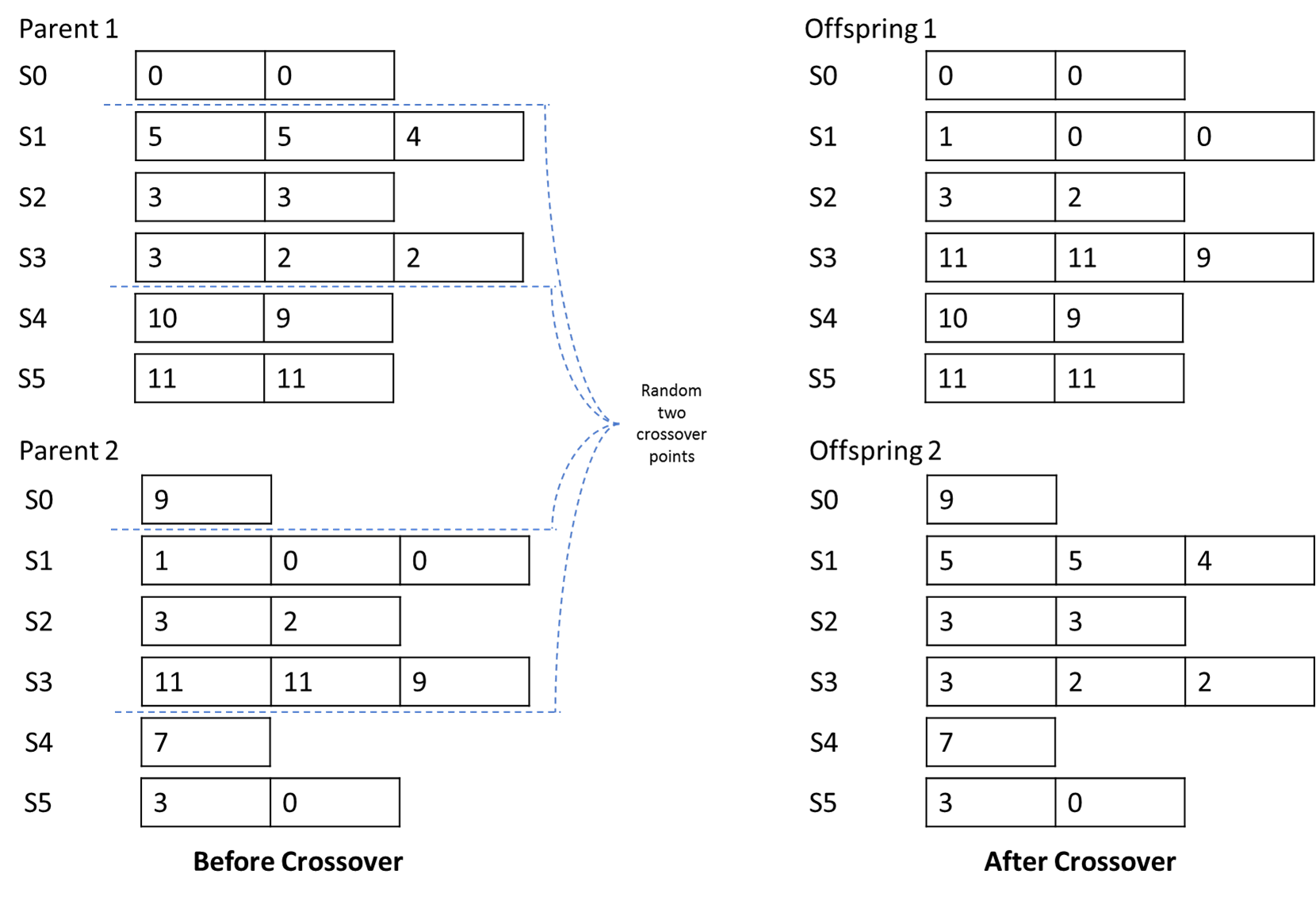}
		\caption{Crossover in the context of our solution.}
		\label{fig:crossover}		
	\end{minipage}
	\hspace{0.1cm}
	\begin{minipage}[t]{1\linewidth}
		\centering
		\includegraphics[width=0.7\linewidth]{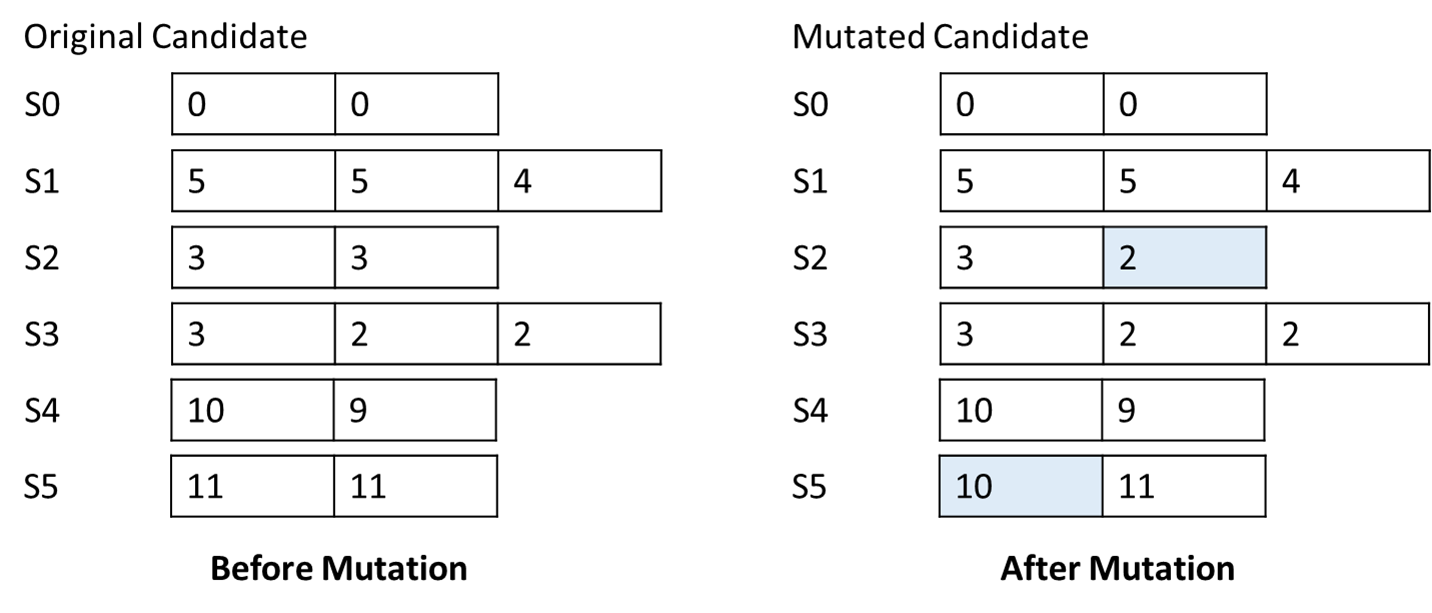}
		\caption{Mutation as performed by our solution.}
		\label{fig:mutation}
	\end{minipage}
	\hspace{0.1cm}
	\begin{minipage}[t]{1\linewidth}
		\centering
		\includegraphics[width=0.7\linewidth]{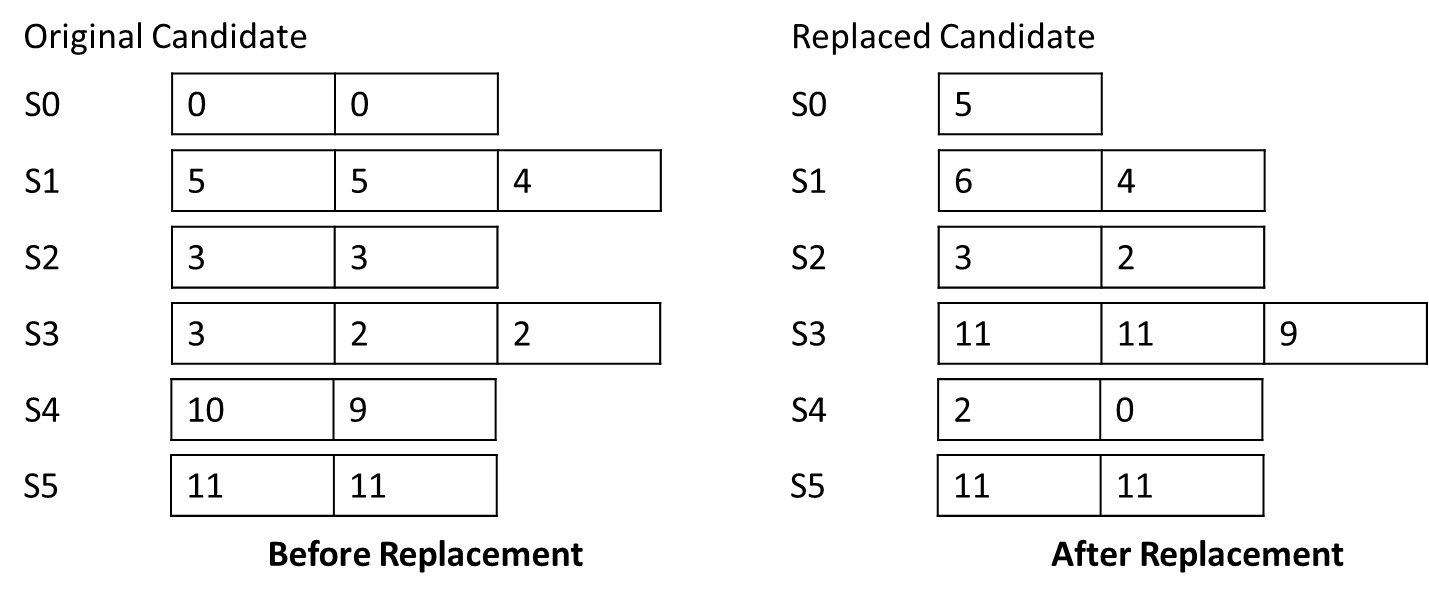}
		\caption{Replacement in our solution.}
		\label{fig:replacement}	
	\end{minipage}
\end{figure*}

\subsubsection{Crossover}
To maintain our assumption regarding to resource provisioning for a service (i.e. resources are provisioned from one cloud in one candidate solution, possibly different clouds in different candidate solutions) and to avoid producing invalid candidate solutions, we exchange chromosomes of services between two candidate solutions. Thus, our crossover is two-point crossover operator equivalent to twice one-point crossovers that is performed on the selected candidate solutions in the generation with a certain probability. An example of candidate solution crossover is shown in Figure \ref{fig:crossover}. .

\subsubsection{Mutation}

In our problem, applying blind random changes for genes of chromosomes in a candidate solution may generate invalid solution that violates application accuracy and performance constraints. Hence, our mutation operator is mutating the candidate solution intelligently by replacing the random gene (VM) in one of its chromosomes with new gene (new VM) that does not violate minimum data processing requirement and has lower provisioning cost. In case of no such new gene is found that meets the requirement, the chromosome of candidate solution is left without mutation, and the other chromosomes of this candidate solution are subjected to mutation based on mutation probability. An example of candidate solution mutation is shown in Figure \ref{fig:mutation}.

\subsubsection{Replacement}

This operator replaces those solutions from the selected candidate solutions whose fitnesses are greater than average fitness based on replacement probability. Each of those solutions is replaced with randomly generated solution if the fitness of the new solution is less than its fitness; otherwise, such solution is retained in the population. It tries twice to find a better solution for each of those solutions, keeping the number of trials at an acceptable level. An example of candidate solution replacement is shown in Figure \ref{fig:replacement}.

\end{document}